\documentclass[twocolumn]{emulateapj}
\usepackage{graphics, enumerate, graphicx, color}
\usepackage{psfig, epsfig}
\usepackage{alltt}
\usepackage{natbib}
\usepackage{epstopdf}
\newcommand{\bearr}{\begin{eqnarray}}
\newcommand{\eearr}{\end{eqnarray}}
\definecolor{titlecol}{rgb}{0,0,1}
\newcommand{\eqnstart}   {\begin{equation}}
\newcommand{\eqnend}     {\end{equation}}

\def\lbol       {$L_{bol}$}
\def\llbol      {$\log L_{bol}$}
\def\mlbol      {L_{bol}}
\def\lbolX      {$L_{bol, X}$}
\def\llbolX     {$\log L_{bol, X}$}
\def\mlbolX     {L_{bol, X}}
\def\ldust      {L_{dust}}

\def\lsum       {$L_{\rm tot, obs}$}
\def\llsum      {$\log L_{\rm tot, obs}$}

\def\ltuv       {$L_{bol,\rm T09}$}
\def\lltuv      {$\log L_{bol,\rm T09}$}
\def\mltuv      {L_{bol,\rm T09}}
\def\KXlo       {14.4}

\def\dKXlolo    {7.7}
\def\dKXlohi    {18.6}
\def\KXhi       {15.2}

\def\dKXhilo    {7.4}
\def\dKXhihi    {16.5}
\def\MBHmed     {5.1 \times 10^8 M_{\odot}}
\def\MBHmedr    {5 \times 10^8 M_{\odot}}
\def\ebv        {$E\left(B-V\right)$}
\def\mebv       {E\left(B-V\right)}

\newcommand{\chisq}{\chi^2}

\newcommand{\lx}{L_X}

\newcommand{\lum}{{\rm erg~s^{-1}}}
\newcommand{\colden}{{\rm cm^{-2}}}

\newcommand{\mbh}{$M_\bullet$}
\newcommand{\mmbh}{M_\bullet}

\slugcomment{Accepted to Ap. J.}

\shorttitle{Obscured GOODS AGN at $z < 1.25$: Slow Growth}
\shortauthors{Simmons et al.}

\begin{document}

%\title{Eddington Ratios of Obscured GOODS AGN}
%\title{Bolometric Luminosities From AGN/Host Morphologies: Obscured GOODS AGN at $z < 1.25$ Are Re-triggered}
%\title{Properties of Obscured GOODS AGN at $z < 1.25$: Re-Triggered Accretion}
\title{Obscured GOODS AGN and Their Host Galaxies at $z < 1.25$: The Slow Black Hole Growth Phase}

\author{B. D. Simmons$^{1, 2}$, J. Van Duyne$^3$, C. M. Urry$^{1, 2}$, E. Treister$^{4,5,}$\altaffilmark{7}, A. M. Koekemoer$^6$, N. A. Grogin$^6$ and the GOODS team}
%\author{B. D. Simmons$^1$, J. Van Duyne$^2$, C. M. Urry$^1$ and the GOODS team}
\affiliation{1 Astronomy Department, Yale University, New Haven, CT 06511, USA}
\affiliation{2 Yale Center For Astronomy \& Astrophysics, Physics Department, Yale University, New Haven, CT 06511, USA}
\affiliation{3 Northrup Grumman, 5500 Canoga Ave, Woodland Hills, CA 91364, USA}
\affiliation{4 Institute for Astronomy, 2680 Woodlawn Drive, University of Hawaii, Honolulu, HI 96822, USA}
\affiliation{5 Universidad de Concepci\'on, Departamento de Astronom\`ia, Casilla 160-C, Concepci\'on, Chile}
\affiliation{6 Space Telescope Science Institute, 3700 San Martin Drive, Baltimore, MD 21218, USA}
\altaffiltext{7}{\emph{Chandra} Fellow.}

\email{brooke.simmons@yale.edu}

\begin{abstract}
We compute black hole masses and bolometric luminosities for 87
obscured AGN in the redshift range $0.25 \leq z \leq 1.25$, selected
from the GOODS deep multi-wavelength survey fields via their X-ray
emission. We fit the optical images and obtain morphological
parameters for the host galaxy, separating the galaxy from its central
point source, thereby obtaining a four-band optical SED for each
active nucleus. We calculate bolometric luminosities for these AGN by
reddening a normalized mean SED of GOODS broad-line AGN to match the
observed central point-source SED of each obscured AGN.  This estimate
of \lbol\ has a smaller spread than simple bolometric corrections to
the X-ray luminosity or direct integration of the observed
multi-wavelength SED, suggesting it is a better measure. We estimate
central black hole masses from the bulge luminosities. The black hole
masses span a wide range, $7 \times 10^6 M_{\odot}$ to $6 \times 10^9
M_{\odot}$; the median black hole mass is $\MBHmedr$. The majority of
these AGN have $L/L_{Edd} \le 0.01$, and we detect no significant
evolution of the mean Eddington ratio to $z = 1.25$. This implies that
the bulk of black hole growth in these obscured AGN must have occurred
at $z \gtrsim 1$ and that we are observing these AGN in a slow- or
no-growth state.
\end{abstract}

% Possible:
% galaxies: active
% galaxies: nuclei
% galaxies: fundamental parameters
% galaxies: Seyfert
% galaxies: bulges
% infrared: galaxies
% X-rays: galaxies
% methods: data analysis

\pagebreak

\section{Introduction}

Accurately estimating the mass and growth of supermassive black holes
(SMBHs) in the centers of massive galaxies is critical to the field of
galaxy formation and evolution.  Calculations of black hole masses
from direct observables have been limited to only a few methods:
stellar kinematics from within the black hole sphere of influence
\citep{genzel97,ghez05}; H$_2$O maser kinematics
\citep{miyoshi95,herrnstein99,greenhill02}; kinematics of central
gaseous disks \citep{ford94,ferrarese99}; and reverberation mapping
\citep{peterson93, peterson00}.
%gas kinematics determined from broad lines in unobscured
%quasars \citep{gebhardt00b,vestergaard02,kelly09}; 
These methods, while powerful and tested, are mostly restricted to
bright galaxies and luminous active galactic nuclei (AGN) in the local
universe.

Far more common are obscured AGN, which by definition have a reddened
(fainter) central point source and often resemble a normal galaxy at
optical and near-infrared wavelengths.  These comprise a large
fraction of all AGN and contribute much of the X-ray background
\citep{ueda03,treisterurry05,treisterurry06,treister04,treister06,treister08,treister09}.
The high luminosity ratio between host galaxy and obscured point
source makes indirect black hole mass (\mbh ) estimates, such as the
$\mmbh-L$ (where $L$ is the galaxy or bulge luminosity) relation
\citep{marconi_hunt03}, much more promising than in quasar hosts, as
well as one of the only methods available given that nuclear emission,
including broad lines, is usually obscured in these sources.

Combining black hole masses with bolometric luminosities gives an
important intrinsic property of AGN, the Eddington ratio, $\lambda
\equiv L/L_{Edd}$, as well as the associated dimensionless accretion
rate, $\dot{m}_{acc} \equiv \dot{M}c^2/L_{Edd}$.  These quantities
indicate how fast a black hole is growing: only accretion approaching
(or exceeding) the Eddington limit leads to an appreciable increase in
black hole mass. From well-studied spectral energy distributions
(SEDs) of unobscured Type~1 quasars
\citep{sanders88,elvis94,richards06}, one can derive bolometric
corrections for AGN with less complete SEDs
\citep{elvis94,fi99,elvis02,marconi04,hopkins07}.  However, these
bright quasars are rare among the larger AGN population (AGN found in
deep \emph{Chandra} exposures are generally $10-100$ times less
luminous) and bolometric corrections for the more common,
lower-luminosity AGN are much less certain.

Detailed study of large samples of AGN and host galaxies has recently
become possible via large multi-wavelength surveys (e.g., GOODS,
COSMOS, ECDFS) that include X-ray and infrared data as well as
high-resolution, deep optical imaging with the Hubble Space Telescope
(\emph{HST}). The unparalleled resolution of the Advanced Camera for
Surveys (ACS) resolves typical AGN host galaxies and point sources out
to $z\sim 1$ \citep[][but also see, \emph{e.g.,} \citeauthor{grogin03}
\citeyear{grogin03} and \citeauthor{schawinski11}
\citeyear{schawinski11} for other \emph{HST}
instruments]{sanchez04,ballo07,pierce07,alonsoherrero08,gabor09}, and
simulations have confirmed the reliability of host and point-source
separation for obscured AGN out to these redshifts \citep{simmons08}.
Here, we utilize the Great Observatories Origins Deep Survey
\citep[GOODS,][]{giavalisco04} X-ray through 24~\micron\ data to
characterize separately the active nuclei and the host galaxies of
a large sample of moderate luminosity, obscured AGN.

The separation of the host galaxy from the AGN point source allows us
to determine simultaneously the SED of the AGN alone and the \mbh\
(from the host galaxy, using the \mbh-$L_{B}$ relation, where $L_{B}$
is the rest-frame $B$-band bulge luminosity).  We can then extrapolate
the AGN luminosity to the far-infrared based on the level of reddening
in the optical point source, and thus estimate the bolometric
luminosity.  Coupled with black hole mass, this allows us to determine
Eddington luminosities and ratios for obscured AGN over the redshift
range of $0.25 < z < 1.25$. We discuss the data and sample selection
in Section \ref{data}. Section \ref{bhmass} describes the black hole
mass estimation, including treatment of time variability of the black
hole-bulge luminosity relation, and Section \ref{lbol} details how we
calculate bolometric luminosities and discusses the reliability of our
methods. In Section \ref{ledd} we discuss the derived Eddington
luminosities and ratios for our sample.

Throughout this paper, we adopt $H_0 = 71$ km/s/Mpc, $\Omega_M =
0.27$$, \Lambda_0 = 0.73$, consistent with the \emph{WMAP} cosmology
\citep{spergel03}.

\section{Data}\label{data}

\subsection{Sample Selection}\label{selection}

In order to reliably extract AGN light from the combined light of an
AGN plus its host galaxy, we require deep observations using an
instrument with a point-spread function (PSF) that is small compared
to the size of the galaxy.  The excellent depth and resolution of the
of the GOODS \emph{HST} observations provide an excellent opportunity
for studying moderate-luminosity AGN ($L_X = 10^{42}-10^{44} \lum$)
out to high redshifts ($z \ge 2)$. The GOODS data include the
space-based Chandra Deep Fields in the X-ray \citep[0.5-8
keV;][]{giacconi02,a03}, four \emph{HST}/ACS filters in the optical
\citep[F435W, $B$; F606W, $V$; F775W, $i$; F850LP,
$z_{850}$;][]{giavalisco04}, all four \emph{Spitzer}/IRAC bands in the
infrared (3.6, 4.5, 5.8, 8.4 $\mu$m), \emph{Spitzer}/MIPS 24~\micron\
\citep[][Dickinson et al., in preparation, Chary et al., in
preparation]{treister06} and the ground-based FLAMINGOS $J$ and $K$
bands and SOFI/ISAAC $JHK$ bands for GOODS-North and South,
respectively.

We define our AGN sample by the following criteria:
\begin{enumerate}
\item \emph{Chandra} X-ray point source matched to an
optically-detected source using a maximum-likelihood method to more
than 99\% confidence \citep{bauer04}.
\item Spectroscopic redshifts \citep{cowie03,wirth04,szokoly04} at $z
\leq 1.25$, so that the data cover the rest-frame B-band.
\item Total absorption-corrected hard X-ray luminosity, $L_X$ (2-8
keV) $ \geq 3\times 10^{42} \lum$ to minimize contamination from
pure starburst galaxies with no AGN \citep{persic04}.

\end{enumerate}

This results in an initial sample of 121 AGN, with 68 from GOODS-North
and 53 from GOODS-South. We apply further criteria to select a
sub-sample of sources with high enough \emph{HST} data quality for
two-dimensional host galaxy fitting.

\begin{figure}
\figurenum{1}
%\epsscale{0.95}
\plotone{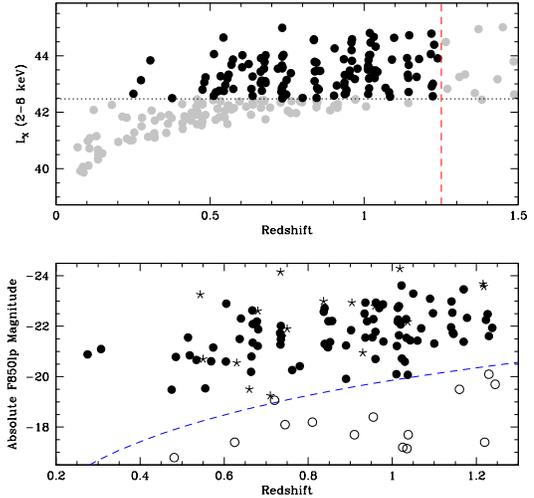}\label{final_cut}
%\plotone{selection_criteria_new.eps}\label{final_cut}
\caption{\emph{Top:} Hard X-ray luminosity vs. redshift for GOODS
X-ray-selected, optical $z_{850}$-detected galaxies with spectroscopic
redshifts. The AGN sample defined by our X-ray and redshift selection
criteria (dashed lines; $L_X > 3\times10^{42}~\lum$, $z < 1.25$) are
black; excluded sources are gray. \emph{Bottom:} Absolute $z_{850}$
host magnitude vs. redshift for the sample AGN in the top panel.  For
accurate morphological fitting, we also require sources with host
$z_{850} < 24$ mag (blue dashed line) and a stellarity index
indicative of an extended source ($<0.85$). All 87 sources meeting
those criteria are shown as black filled circles. Faint sources with
host $z_{850} > 24$ are shown as open circles; point-like sources with
host $z_{850} < 24$ are shown as stars.  }
\end{figure}

To separate a source's central (optical) AGN light from its host
galaxy light using two-dimensional parametric fitting techniques
requires high-resolution data and relatively high signal-to-noise for
each source ($>5$ per pixel). To ensure the highest quality fits, we
include only sources with host $z_{850} \leq 24$~mag (AB), and with a
stellarity parameter \citep[estimated by the SExtractor parameter,
{\tt CLASS\_STAR};][]{sextractor} less than 0.85, where ${\tt
CLASS\_STAR}=1$ is a pure point source.  Following this cut, we retain
90 AGN (51 from GOODS-N, 39 from GOODS-S), of which 9 are broad-line
objects. Figure \ref{final_cut} shows the distributions of $L_X$,
$M_{850lp}$, and $z$ for X-ray sources with $z < 1.25$ in the GOODS
fields. We further exclude three sources for which we were unable to
isolate the bulge flux. Two have extremely irregular and disrupted
optical morphology, making morphological descriptions in terms of
bulges and disks meaningless, and one has a marginal ${\tt
CLASS\_STAR}=0.845$; all three are among the lowest optical
luminosity, with a mean $M_B = -19.05$, nearly an order of magnitude
below the next lowest AGN host ($M_B = -20.2$; see
Figure~\ref{final_cut}).  The remaining 87 AGN+hosts constitute our
final sample.

\subsection{Point Source-Host Galaxy Decomposition}\label{galfit}

\begin{figure*}
\figurenum{2} %\epsscale{1.0} 
%\psfrag{xlabel}{$\log \nu$ [Hz]}
%\psfrag{ylabel}{$\log \nu L_{\nu}$ [erg/s]}
%\plotone{f2.eps}\label{galfits}
%\plotone{f2.png}\label{galfits}
\plotone{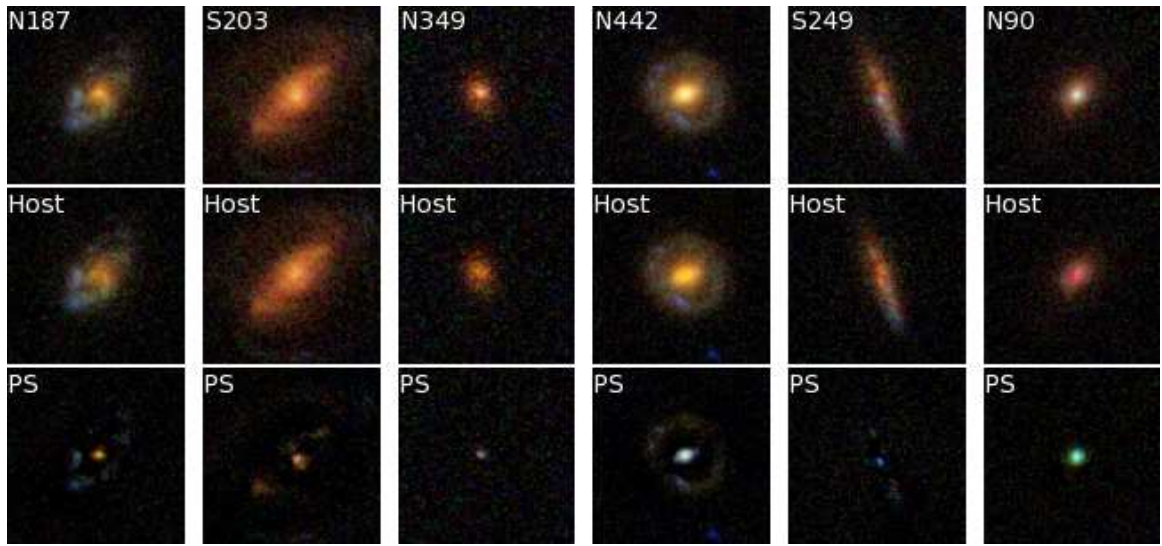}\label{galfits}
\caption{$BVIz$ composite images of six of the sources in our
sample. Each source, marked with its ID from \citet{a03}, is fit in
each band with a host galaxy composed of a bulge and disk, and a
central point-source. For each source, the image marked ``Host'' shows
the point-source-subtracted galaxy. The image marked ``PS'' shows the
residuals when the smooth host galaxy fit is subtracted, leaving
behind the central nucleus and extended detailed features of the host
galaxy. Hosts and point sources vary in color and luminosity. (SEDs of
these sources are shown in Figure~\ref{seds}.)}
\end{figure*}

We perform morphological decomposition of the GOODS AGN+host galaxies
using the 2-D fitting routine {\tt GALFIT} \citep{peng02}.  This routine
allows for simultaneous fitting of one or more host galaxy components
along with a central point source. We used the GOODS ACS
images that were processed using the STScI Multidrizzle
algorithm \citep{multidrizzle}, improving the native resolution to
0.03 arcsec/pixel. 

We fit each source independently in each ACS band ($B$, $V$, $I$, and
$z_{850}$) with a three-component fit: a deVaucouleur bulge + an
exponential disk + a nuclear point source, the latter of which is
modeled by a noiseless, analytical PSF based on analysis of dozens of
real stars in the GOODS fields, created independently for each band
using the IRAF package {\tt daophot}. This PSF creation method
minimizes the effects of potential eccentricities of any one star on
the field (such as color or excess noise), while still accounting for
any possible deviations within our drizzled data from a purely
analytical PSF created from a package such as {\sc tiny tim}
\citep{krist93}.

Each galaxy was initially fit using an automated program
\citep[described in further detail in][hereafter S08]{simmons08} that
uses SExtractor catalog values as initial parameter guesses. The
primary goal of the initial fit is to fix the central positions of the
host and point source and to calculate initial estimates of the fit
parameters.  Each galaxy was then fit with {\tt GALFIT} by hand to achieve
the best possible fit, assessed using the $\chi^2$ goodness-of-fit
parameter and examination of fit residuals. We then calculate the
rest-frame $B$-band bulge-to-total ratios and point-source and host
galaxy luminosities, using the InterRest interpolation code from
\citet{taylor09}; see Table~1 for total, point source, and host galaxy
magnitudes and associated errors. Separated hosts and point sources
for six objects in our sample are shown in Figure~\ref{galfits}.

Reliable calculations of AGN black hole mass and bolometric luminosity
require accurate measurements of each source's host galaxy-to-point
source luminosity and bulge-to-total ratios, which are determined by
our morphological fitting. S08 simulated over 50,000 AGN host galaxies
in order to assess the limits of the morphological fitting. The
simulations show that the determination of $L_{host}/L_{PS}$ is
generally reliable to the flux and redshift limits of our
sample. Brighter point sources introduce greater uncertainty into the
recovered parameters, especially the host galaxy magnitude and
bulge-to-disk ratio (S08). We add these uncertainties in quadrature
with the fit errors, which in turn affects the uncertainties in our
black hole masses and Eddington ratios.

For samples like ours, automated fitting of AGN and hosts recovers at
least 90\% of central point sources (S08); our recovery fraction
should be higher due to the fact that we fit each source
individually. S08 also find a spurious point-source detection rate of
approximately 12\% in a single-band sample with equal numbers of
bulge- and disk-dominated sources. However, this is an upper limit for
our sample, since the chance of spurious detections or unrecovered
point sources in multiple bands for a single source is extremely
small.  Statistically, the probability of a missed point source in 2
(3, 4) bands is 1\% (0.1\%, 0.01\%) for one source. The probability of
a spurious detection in 2 (3, 4) bands for a source with equal
contributions from a bulge and disk is 1.4\% (0.2\%, 0.02\%), though
this calculation varies with the morphology of each source. Based on
the number of point sources detected in our sample and each source's
individual morphology (following S08), we expect that we miss $\lesssim 3$
point source detections within the sample and have spurious detections
for $\sim 2$ sources.

Rest-frame B-band properties of each host galaxy and central point
source are given in Table~1.

\section{Black Hole Mass Estimation}\label{bhmass}

\begin{figure}
\figurenum{3}
\plotone{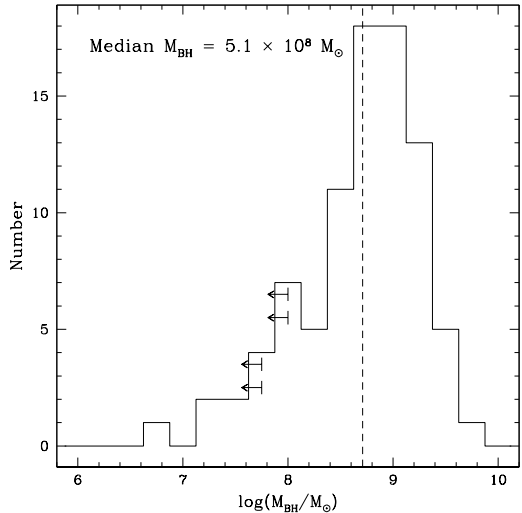}\label{bhm_hist}
%\plotone{BHstats.eps}\label{bhm_hist}
\caption{Histogram of estimated black hole masses, using bulge
luminosities from morphological fitting and Equation
\ref{ML_equation}, adapted from \citet{marconi_hunt03}.  The dashed
line shows the median value of $\mmbh = \MBHmed$. Arrows indicate the
four sources with upper limits to the black hole mass.}
\end{figure}

Direct kinematic measurements of black hole mass are not possible for
the majority of our sample because most of the AGN are
obscured. Instead we use black hole-bulge relations to estimate the
black hole masses. We use the point-source-subtracted bulge luminosity
to calculate \mbh\ from rest-frame $B$-band absolute magnitudes of the
host galaxy bulge using the $\mmbh-L_B$ relation from Eq. 19 of
\citet{ff05} as converted from \citet{marconi_hunt03}:
\begin{equation}\label{ML_equation}
\rm \log(\mmbh) = (8.37 \pm 0.11) - (0.419\pm0.085)(B_B + 20.0),
\end{equation}
where $B_B$ is the rest-frame B-band absolute magnitude of the galaxy
bulge. Rest-frame $B$ bulge-only absolute magnitudes range from
$-16.32 > B_B > -23.27$, with an average error of $\sigma_B = 0.13$
(determined from {\tt GALFIT} uncertainties added in quadrature to
systematic uncertainties from S08). Within our sample, we find bulge
fractions ranging from $<5$\% to 100\%, with a median bulge-to-total
ratio of $0.54$. For the four cases where a bulge is undetected in the
rest-frame $B$, we follow S08 in assuming a maximum of 5\% bulge
contribution, and consider those black hole masses upper limits.

The mass-luminosity relation of Eq.~\ref{ML_equation} is based on
measurements in the local universe.  \citet{treu04} have shown an
evolution in the relation out to redshift $z=0.3$: the rest-frame
$B$-band mass-to-light ratio of the bulge decreases with redshift due
to the fact that a higher-redshift bulge typically has a younger
stellar population than a bulge at $z=0$. Subsequent work on different
samples and using different methods
\citep{borys05,alexander08,bluck11} finds that black hole masses
seem to lag behind bulges at even higher redshift, implying that black
hole masses calculated from bulge properties are overestimated by
approximately a factor of 3 at $z \sim 1$.

However, several other studies
\citep{woo05,peng06,woo08,jahnke09,decarli10b,merloni10} report a
decrease in the intrinsic bulge stellar mass-black hole mass relation
with redshift, in the sense that bulge mass appears to evolve faster
than black hole mass within our redshift range, leading to an
evolution in the bulge-black hole mass relation in the opposite
direction. Some of the conflicting results may be explained by
differences resulting from comparing black hole masses to different
quantities (stellar mass versus bulge luminosity, for example), but
the picture is not yet clear. Black hole masses calculated at $z \sim
1$ may change by a factor of approximately 3 in either direction and
still be consistent with previous studies.

Going forward, we quote the unevolved masses, calculated via a Monte
Carlo method in order to account for uncertainties in host luminosity
and bulge-to-total ratio (both determined from S08) as well as the
intrinsic scatter noted in Equation~\ref{ML_equation} and an
additional uncertainty of $\pm 0.47$~dex to encompass the uncertainty
in the evolution of Eq.~\ref{ML_equation} to the redshifts relevant to
our sample. We generate $10^5$ data points for each of our 87 sources,
with uncertainties folded into each step to calculate black hole
masses; the reported \mbh\ values and errors represent the peak and
$\sigma$ values of the Monte Carlo distribution for each object. This
method typically results in SMBH mass uncertainties of approximately
$\pm 0.5$~dex in log-\mbh . Using the same Monte Carlo method, we
propagate these errors through to our calculation of Eddington ratios
(discussed in Section~\ref{ledd}).

The resulting black hole masses are shown in Figure \ref{bhm_hist} and
presented in Table~1.  The majority (94\%) are between
$5.0\times10^7 < \mmbh < 5.5\times10^9~\rm M_\odot$, and the rest have
masses (or upper limits) down to $\mmbh = 6.7\times10^6~\rm M_\odot$.
%(Figure~\ref{mass_estimate}).  
The median mass, $\mmbh = 5.1\pm4.7\times10^8~\rm M_\odot$, is
indicated on Figure \ref{bhm_hist} as a dashed line.

%% As a sanity check on the accuracy of the black hole masses, we checked
%% the three broad-line AGN in GOODS-South with available spectral line
%% widths \citep[IDs 62, 68, 117, in][]{szokoly04}.  Black hole masses
%% are calculated using the \mbh$-\sigma$ relation as derived from the
%% FWHM of [O~III]$\lambda~5007$ \citep{nelson00, grupe_mathur04,
%% boroson_2003,boroson_2005, gh07b}.  Converting line width to velocity,
%% we derive mass estimates that are within 10-18\% of the masses
%% estimated from the host galaxy luminosity. Although a statistically
%% small sample, the \mbh\ values are accurate to within the $1\sigma$
%% uncertainties of both methods.

\section{Bolometric Luminosity Calculations}\label{lbol}

\begin{figure*}
\figurenum{4}
\plottwo{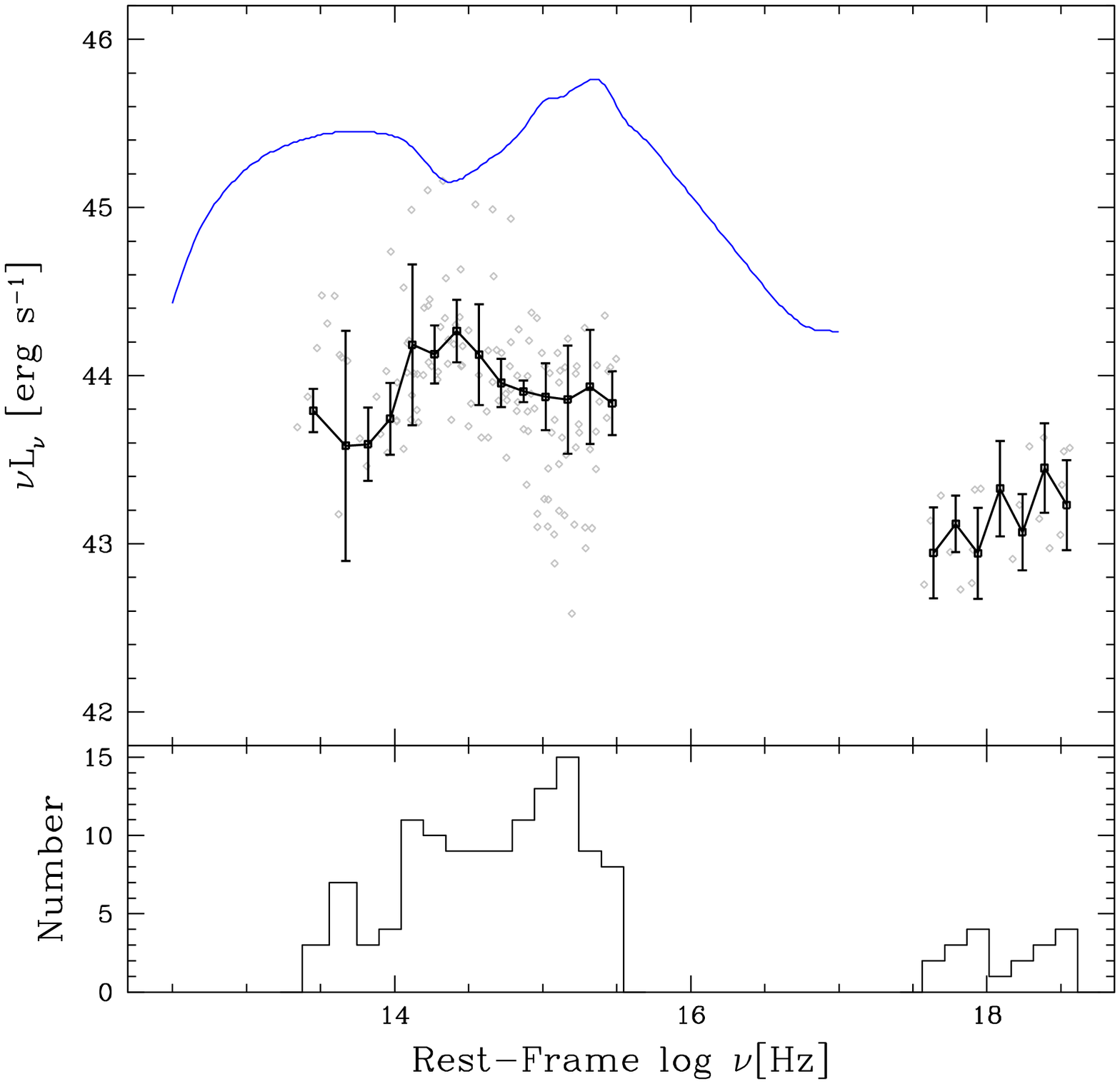}{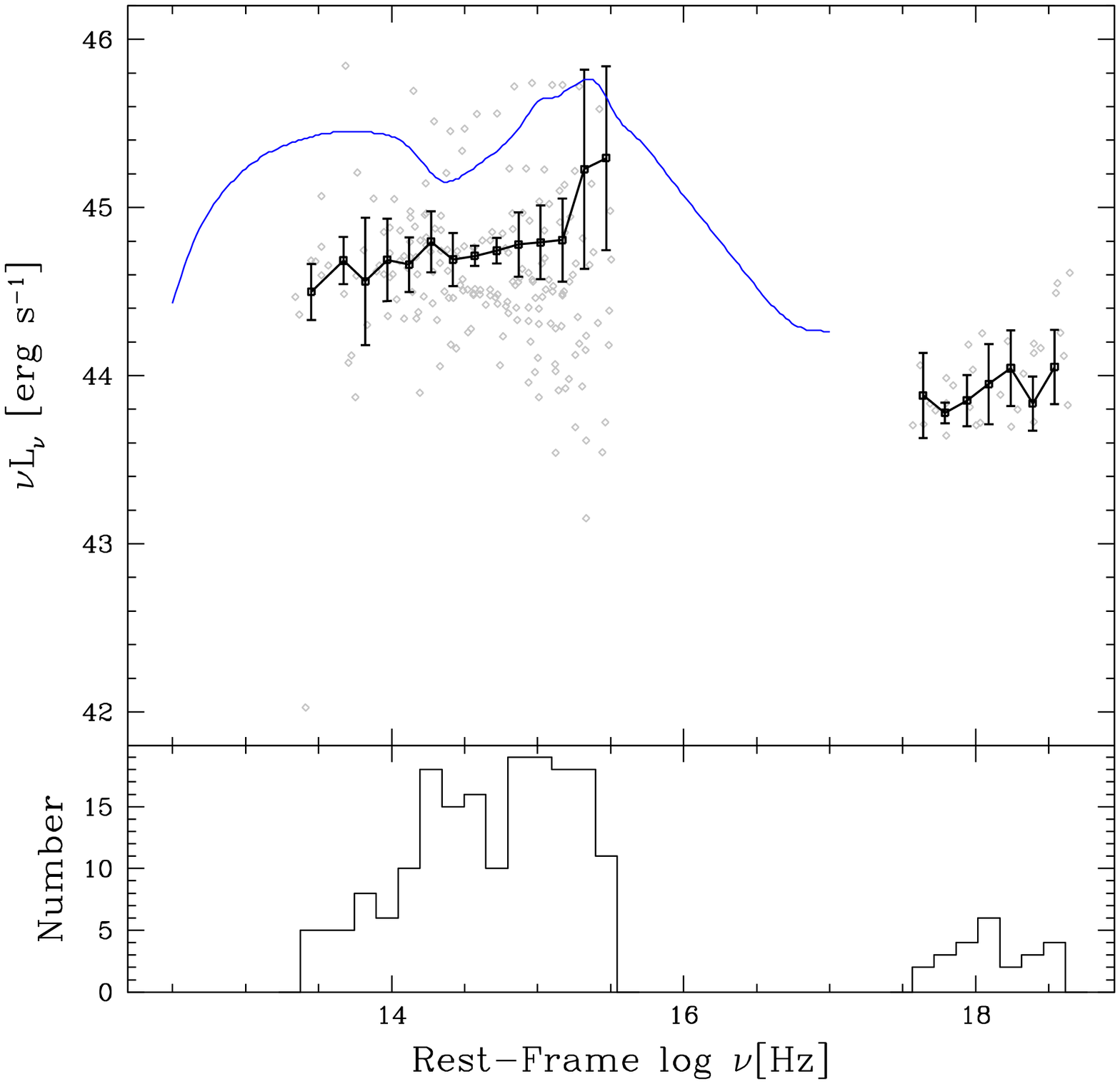}\label{type1_hilo}
%\plottwo{Xray_BC_corrections_paper_lolum.ps}{Xray_BC_corrections_paper_hilum.ps}\label{type1_hilo}
\caption{Weighted average SEDs of broad-line AGN in the GOODS fields
(filled circles). The left panel shows 32 low-luminosity ($43.0 < \rm
\log L_X < 44.0$) AGN, and the right panel shows 26
moderate-luminosity ($44.0 < \rm \log L_X < 44.9$) AGN.  The bottom
histogram in each panel shows the number of AGN contributing to the
median, as a function of frequency. For comparison, the average
spectrum of more luminous SDSS quasars \citep[$L_X >
10^{45}~\lum$;][]{richards06} is shown as a blue dashed line. Our SEDs
are very roughly consistent with the Richards et al. spectrum when
normalized at $~\sim 1$ keV, but the ``big blue bump'' feature at
$\sim2500\AA$ is significantly weaker in lower luminosity AGN. Note
that some lower-luminosity objects used to make the composite SED in
the left panel show signs of a K-giant stellar population in their SEDs. }
\end{figure*}

In order to study the properties of AGN, we need to know the total
power they emit. Bolometric luminosities of AGN are thus important
quantities, but they are also difficult to calculate directly for many
populations of AGN. Owing to our selection criteria, the SEDs of our
87 sources are typically dominated by galaxy light at optical and
near-IR wavelengths, so we do not expect that naively summing the
luminosities of our sources at all wavelengths will provide a reliable
estimate of total AGN power.

We therefore consider two alternative approaches: (1) an average
bolometric correction to the X-ray luminosity based on unobscured
broad-line AGN (\S~\ref{KX}), and (2) individual corrections based on
fitted reddening of our main AGN sample (\S~\ref{dust_lum}).

\subsection{X-ray Bolometric Correction}\label{KX}

Bolometric luminosities are commonly estimated using an X-ray
bolometric correction, $K_X$, where $\mlbolX = K_X L_X$.  Values of
$K_X$ have been derived from well-defined SEDs of optically selected,
powerful quasars and range from $K_X \sim 10$ \citep{elvis94} to $K_X
\approx 33$ \citep{fi99} or $K_X > 50$ \citep{elvis02}. The $K$ factor
is luminosity- and wavelength-dependent
\citep{marconi04,lafranca05,treisterurry05}.

The bulk of our sample lies more than an order of magnitude below the
average $L_X$ of mean quasar SEDs such as those computed from the
Sloan Digital Sky Survey \citep{richards06}.  This suggests the $K_X$
values from the literature may not accurately represent our sample.
Since the GOODS data cover nearly 5 decades of wavelength, we
calculate our own $K_X$ value and compare to previously derived $K_X$.

First, we develop an SED for unobscured, broad-line AGN by averaging
the 58 GOODS sources with clear broad lines and hence low reddening
\citep{barger03,szokoly04,cowie04}, as well as $N_H < 10^{21}~\colden$
\citep{bauer04}. We also impose the criterion $\lx > 10^{43}~\lum$ to
minimize host galaxy contamination in the optical and near-IR.  (Three
of these sources are in our sample of 87 AGN.) Even under these
conditions, 12 of the selected broad-line AGN show some signs of a
K-giant stellar population in their SEDs \citep{barger03,mainieri04},
particularly at lower $L_X$.  We then split the unobscured sample into
two groups, less than and greater than $L_X = 10^{44}~\lum$, to
investigate the dependency of $K_X$ on $L_X$.  The median,
absorption-corrected, 2-8 keV X-ray luminosity for each group is
$2.4\times10^{43}$ and $1.4\times10^{44}~\lum$, respectively,
consisting of 32 and 26 AGN, respectively.

Figure \ref{type1_hilo} shows the rest-frame, error-weighted geometric
mean of our intermediate-luminosity AGN in frequency bins 0.15 dex in
width.  Given the wide redshift range, not all wavelength bins include
the same number of photometric bands for each unique object.  In
addition to weighting by photometric error in each bin, we also weight
by the fraction of AGN in each band with respect to the total sample
number (shown in the bottom panel of Figure \ref{type1_hilo}).

For comparison, we also plot in Figure \ref{type1_hilo} the mean SDSS
quasar SED from \citet{richards06}.  The ``big blue bump'' in both
GOODS AGN samples (at $\sim2500\AA$) is considerably weaker than in
SDSS quasars, relative to the mean luminosity at 1~\micron.  This
suggests that using an SDSS-derived $K_X$ estimate would overestimate
the bolometric luminosity in our AGN sample.

The integrated luminosities of the averaged broad-line AGN source SEDs
are calculated by summing the 2-30~keV rest-frame $L_X$, the 1250
\AA-8~\micron\ broad-band luminosities and the integrated UV-to-soft
X-ray assuming $L_\nu \propto \nu ^{1.41}$.  This results in
integrated luminosities and bolometric corrections of
$L_{bol,BL}(8~\micron-30\rm keV) = 4.1 \times 10^{44}~\lum$ and $K_X =
\KXlo_{-\dKXlolo}^{+\dKXlohi}$ for the lower-luminosity broad-line
AGN, and $L_{bol,BL}(8~\micron-30\rm keV) = 2.6 \times 10^{45}~\lum$
and $K_X = \KXhi_{-\dKXhilo}^{+\dKXhihi}$ for the higher-luminosity
broad-line AGN. The $K_X$ value in the lower-luminosity broad-line AGN
may be slightly high due to host contamination of the optical and
near-infrared SED (we estimate this contamination increases $K_X$ in
the lower luminosity sample by $\approx 20$\%).

\subsection{Point-Source Reddening / Dust Luminosity Method}\label{dust_lum}

\begin{figure*}
\figurenum{5} %\epsscale{1.0} 
%\psfrag{xlabel}{$\log \nu$ [Hz]}
%\psfrag{ylabel}{$\log \nu L_{\nu}$ [erg/s]}
\plotone{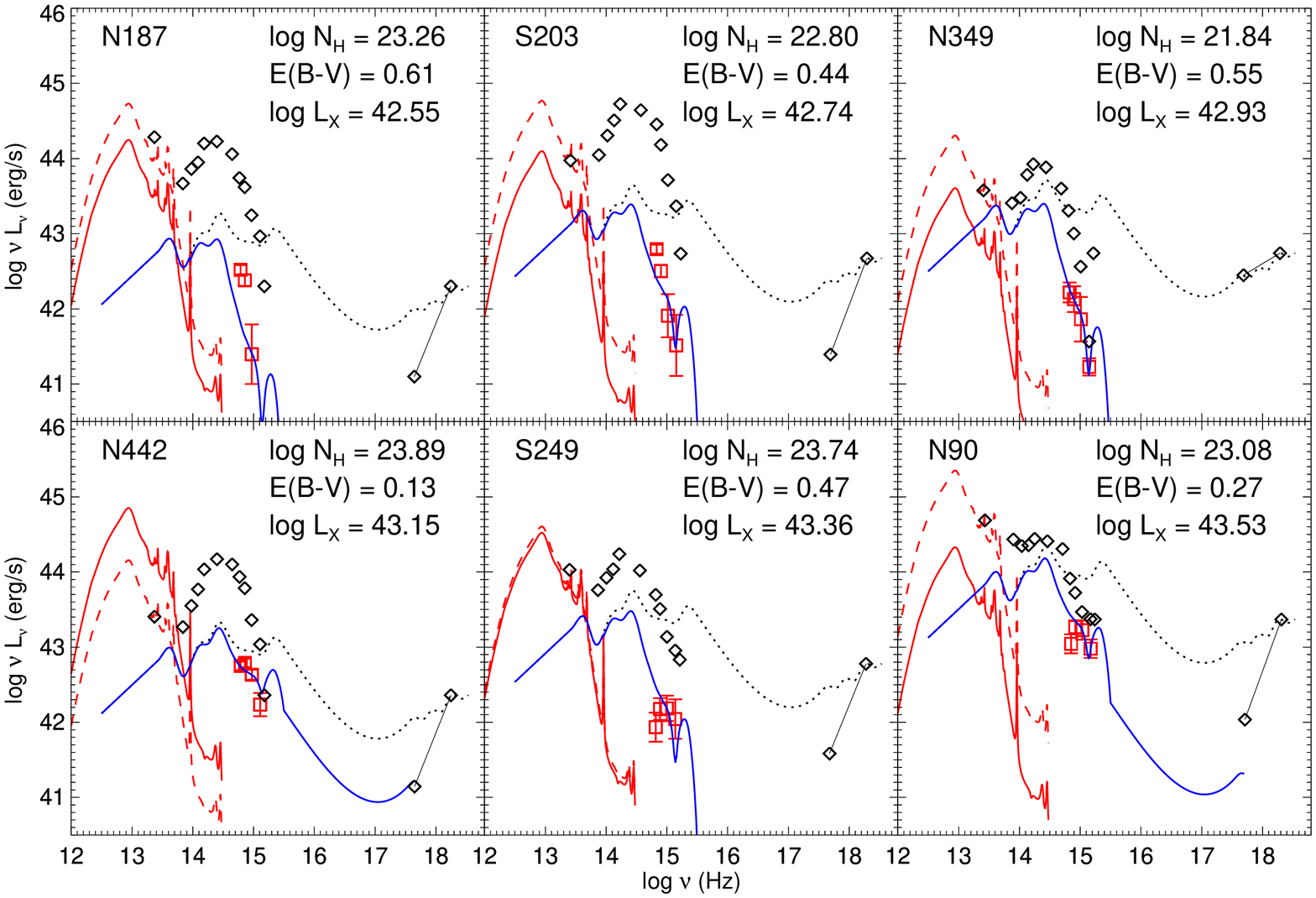}\label{seds}
%\plotone{bhmass_seds_sdss_paper_grid_page0.ps}\label{seds}
\caption{Broad-band SEDs of AGN plus host galaxy (\emph{open
diamonds}) and point source alone (\emph{red squares}).  The mean AGN
SED, calculated from broad-line AGN within the GOODS fields, is shown
as a \emph{dotted line}, normalized to the AGN hard X-ray luminosity.
This template was reddened and fit through the extracted point source
values, resulting in the \emph{blue solid curve}.  The far-IR dust
curve (\emph{solid red line}) is calculated from the derived \ebv\
according to \citet{draine07}. Shown for comparison is the same
emissivity normalized to the maximum $L_{MIR}$ from broad-band flux
(\emph{dashed red line}). In most cases, the calculated (solid) curve
lies under the maximum (dashed) curve, \emph{i.e.,} the calculated
infrared luminosities do not exceed the observed luminosities. In the
rare event that the solid line exceeds the observed infrared flux
(N442), it is usually by a small factor ($\times 2$), commensurate
with AGN variability on year-long timescales
\citep{sarajedini06}. (Images of these sources are shown in
Figure~\ref{galfits}.)}
\end{figure*}

We can derive the observed point-source SED in the four ACS bands
using the results from morphological fitting. In some sources,
typically in the ACS $B$ (and possibly $V$) bands of higher-redshift
sources, no point source is detected at all; in these cases, we
calculate an upper limit for these bands based on the residual
background flux.  Figure~\ref{seds} shows six examples of total
broad-band SEDs (diamonds) from 24~\micron\ to U-band, plus X-ray,
and the point-source optical SEDs as red squares.

We redden an unobscured AGN template to fit the extracted point
source luminosities. We use the two averaged AGN SEDs calculated in
the previous section, choosing which template to use for each source
based on the source's X-ray luminosity. First, we normalize the
template to the absorption-corrected, hard X-ray luminosity of each
source. After this normalization, the mean SED exceeds the observed
rest-frame broad-band MIR luminosity by more than the uncertainty of
the mean SED for only eight of our 87 sources. This indicates that the
choice of template SEDs is appropriate for our sample.\footnote{If we
were to instead use the \citet{richards06} SDSS quasar template, the
template would exceed the MIR flux in 68\% of sources.} In the rare
case where the SED exceeds the observed MIR luminosity after
normalizing to the hard X-ray luminosity, the mean SED is
re-normalized to be consistent with the observed MIR luminosity.

We then fit the normalized spectrum to the optical point-source SED,
applying the reddening curve of \citet{cardelli89} for a Milky Way ISM
($R_V = 3.1$) over the wavelength range of $0.1 < \lambda <
3.5~\micron$ and using a $\chisq$-minimization algorithm to calculate
the reddening value of the nuclear emission. For sources with point
source detections in three or four filters ($N_{pt} \geq 3$), the
reddening average is $\left<\mebv \right> = 0.47 \pm 0.37$. This
increases to $\left<\mebv \right>=0.78 \pm 0.54$ for AGN with solid
detections in only two bands.  We note that point source SEDs with
significantly different colors from the host galaxy SED are frequent
enough in our sample that attempting to determine the reddening value
from the combined AGN+host SED would be considerably less
reliable. This is consistent with previous simulations and
observations \citep[\emph{e.g.},][]{pierce10a,pierce10b}.

The independent determination of \ebv\ from reddening and $N_H$ from
X-ray absorption allows us to estimate the dust-to-gas ratio along the
line of sight (Section \ref{dust2gas}). We can then estimate the total
re-radiated IR luminosity of the material that is absorbing optical
and UV photons from the point source, using the dust emissivity models
of \citet{draine07} and the dust-to-gas ratios to convert the
emissivity profiles to luminosity, $\ldust$.

This calculation requires two assumptions: (1) the ambient radiation
field strength, $U$, near the dust, and (2) the physical line-of-sight
path length, $l$, through the AGN and host galaxy.  We select the
\citet{draine07} model that has a range of $U = (1-1\times 10^5) U_0$,
where $U = \int_{0.09}^{8~\micron} 4\pi J_\lambda = 4.34\times
10^{-2}$ erg~cm$^{-2}$~s$^{-1}$ \citep{mathis83}, to roughly reproduce
the combination of AGN radiation field strength and the ambient galaxy
radiation field.  We estimate the path length to be $l = R \cos i + H
\sin i$, where $R$ is half of the physical galaxy semi-major axis
(determined from the angular size and distance scale), and $i$ is the
estimated inclination.  For galaxies with no detected disk to indicate
orientation, we assume $i=0$.

We find that the calculated $\ldust$ does not exceed the observed
$24~\micron$ luminosity in the majority (83\%) of cases. Of the 15
sources with excess calculated $\ldust$ compared to the observed SED,
only one is inconsistent with AGN variability observed for a similar
sample on year-long timescales \citep{sarajedini06}. That source,
S160, has an uncharacteristically low $24~\micron$ luminosity compared
to the rest of our sample. We conclude that the selected parameters of
$U$ and $l$ do not greatly over- or underestimate the reprocessed AGN
dust emission of our sample. However, the assumption that reddening of
the optical point source is due entirely to dust from an AGN torus
means that the calculated $\ldust$ should be considered an upper limit.

\subsection{Comparison of Bolometric Luminosity Methods}\label{lbol_compare}

\begin{figure}
\figurenum{6}
%\epsscale{1.0}
\plotone{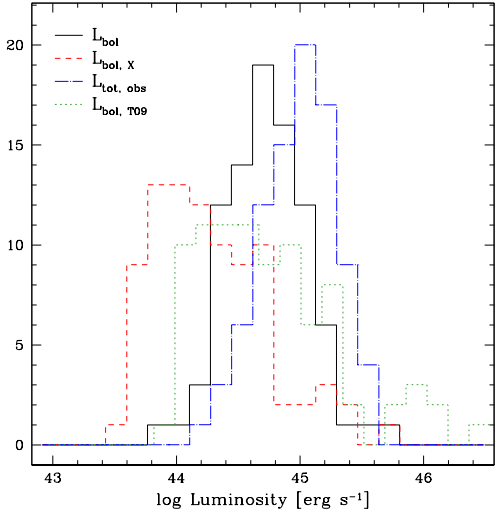}\label{bol_hists}
%\plotone{Lbol_hists.eps}\label{bol_methods}
\caption{Histograms of bolometric luminosities calculated four ways:
  (1) using our SED-reddening fit method from Section \ref{dust_lum},
  \lbol\ (black solid line); (2) using our empirical bolometric
  correction to the X-ray luminosity, \lbolX\ (red, dashed line); (3)
  using a simple sum of all the observed light for each source, \lsum\
  (blue, dot-dashed line); and (4) using the model-based,
  luminosity-dependent correction of \citet{treister09}, \ltuv\
  (green, dotted line). The first method has the smallest scatter and
  agrees well with the successful population synthesis model of
  Treister et al.; the second method underestimates the luminosity,
  probably because a significant fraction of the X-rays are absorbed;
  and the third method overestimates \lbol\ because of a significant
  contribution from the host galaxy. A Kolmogorov-Smirnov (K-S) test
  indicates a 6.3\% chance that the distributions of \lbol\ and \ltuv\
  are from the same parent distribution, whereas there is less than a
  0.001\% chance that the \lbol\ distribution is consistent with
  either the \lbolX\ or \lsum\ distributions.}
\end{figure}

\begin{figure*}
\figurenum{7}
\epsscale{0.8}
\plotone{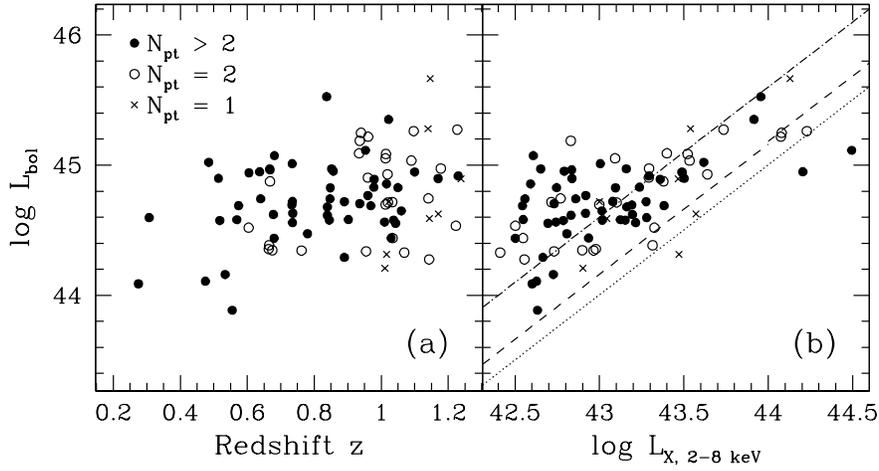}\label{bol_z_X}
%\plotone{Lbolrat.eps}\label{bol_methods}
\caption{Calculated bolometric luminosities and comparison of
methods. Panels show: (a) Bolometric luminosity (\lbol ), as described
in Section \ref{dust_lum} (method 1 in Figure \ref{bol_hists}), versus
redshift. Solid points indicate objects with nuclear point source
detections in at least 3 optical bands ($N_{pt} > 2$); open circles
show sources with $N_{pt} = 2$; crosses show sources with $N_{pt} =
1$, which have highly uncertain values of \ebv\ and thus \lbol . We
detect no trend in \lbol\ with redshift. (b) Bolometric luminosity
versus absorption-corrected hard X-ray luminosity. Fixed bolometric
corrections of $\mlbol = 10 \times L_{X}$ \citep[dotted
line;][]{elvis94} or $\mlbol \approx 15 \times L_{X}$ (dashed line;
Section \ref{KX}) underestimate \lbol . The peak of the $\mlbol /
L_{X}$ distribution ($\left(\mlbol / L_{X}\right)_{\rm peak} = 39.5$)
is shown as a dot-dashed line and is likely higher than corrections
based on unobscured AGN because of significant X-ray absorption in
this sample. }
\end{figure*}

Figure~\ref{bol_hists} shows a comparison between the bolometric
luminosities calculated four ways: (1) by integrating the complete AGN
SED we constructed for each object as described in Section
\ref{dust_lum}, \lbol ; (2) by estimating an X-ray correction using
the mean broad-line AGN SEDs calculated in Section~\ref{KX}, \lbolX ;
(3) by summing the source luminosity in all observed bands, \lsum ;
and (4) by using model-derived, luminosity-dependent bolometric
corrections to the X-ray luminosity \citep[\ltuv ;][]{treister09}.

For the relatively small number of objects (6) in our sample with $L_X
> 10^{44}~\lum$, using our point-source fitting method to calculate
\lbol\ has close to the same result as simply adding up the total
luminosity in the source SED. However, the rest of our sources are
dominated by the host galaxy in the optical and infrared wavelengths,
leading to a significant overestimation of AGN-only luminosity with a
simple summation method. For our entire sample, the mean value of
\lsum\ is too large by approximately 80\% compared to the mean \lbol\
computed using our point-source fitting method.

In contrast, the simple bolometric correction computed in Section
\ref{KX} is too low by 66\% on average. This is unsurprising, because
our correction is derived from an observed SED that lacks FIR data at
longer wavelengths than $24~\micron$ (observed), which can contribute
significantly to the observed total luminosity. This is exacerbated
when applying a correction to objects like those in our main sample,
whose X-ray luminosity has been subject to circumnuclear
absorption. Using the \citet{draine03} dust models to estimate this
FIR component in the broad-line AGN increases the value of the
correction from Section \ref{KX} to a value consistent with
\citet{elvis02}, which increases the corrected luminosities, \lbolX ,
to a mean value more consistent with the mean \lbol , but with a
larger scatter ($\sigma$ for $\log \mlbolX = 0.46$ compared to 0.33
for $\log \mlbol$).
%If we instead use the individually calculated values of \lbol\ to
%create a bolometric-X-ray correction for each object, we find the peak
%and $\sigma$ of the distribution to be $K_{X,\rm PS} =
%39.5^{+40.4}_{-20.0}$. 

Both \lbolX\ and \lsum\ are calculated using observed properties. We
assess the accuracy of our \lbol\ calculation, which combines
observations with model-dependent parameters, by comparing to an
independent theoretical model. Specifically, we use the
luminosity-dependent bolometric corrections calculated in
\citet{treister09} for model AGN SEDs with a range of X-ray
luminosities, column densities, and orientations
\citep{treister06}. Using the absorption-corrected X-ray luminosity of
each object in our main sample to predict a bolometric correction
yields a bolometric luminosity for each object, \ltuv . 

Encouragingly, the mean values of \ltuv\ and \lbol\ are in good
agreement ($\log \left<\mltuv\right> = 44.75 \pm 0.55$ and $\log
\left<\mlbol\right> = 44.74 \pm 0.33$), with the distribution of
$\mltuv / \mlbol$ having a peak and width ($\sigma$) of $1.0 \pm 0.8$. This
indicates that the prediction of the \citet{treister09} model
generally agrees well with calculations of \lbol\ for individual
objects. Such strong agreement for two completely independent ways of
calculating bolometric luminosities implies that the observed
reddening of the central point source is dominated by absorption
processes occurring within the circumnuclear region.

Individual-source bolometric luminosities for each of the four methods
are presented in Table~2. Note that, compared to \lbol\ (method 1),
\lbolX\ (method 2) underestimates the bolometric luminosity and \lsum\
(method 3) overestimates it. Both \lbolX\ and \ltuv\ (method 4)
distributions have more scatter, which suggests the point-source
fitting method (method 1) may be superior. We use this value, \lbol ,
in the subsequent analysis.

Figure \ref{bol_z_X} shows the distribution of \lbol\ with redshift,
as well as the ratio of \lbol\ to the observed $L_X$. We detect no
trend in \lbol\ with redshift, but the objects with lower $L_X$ appear
to have a higher scatter in \lbol\ compared to objects with higher
$L_X$. Interestingly, values of \lbol\ calculated using point-source
luminosities in only 1 or 2 optical bands, while individually more
uncertain than values calculated with 3 or 4 bands of optical
point-source data, are not outliers in the overall distribution of
\lbol . This suggests that a sample of AGN with \lbol\ calculated as in
section \ref{dust_lum}, but with only one optical band used to
determine \ebv , may be reliable as a sample even if the individual
values have high uncertainty.

\subsection{Dust-to-Gas Ratios}\label{dust2gas}

\begin{figure}
\figurenum{8}
%\epsscale{1.0}
\plotone{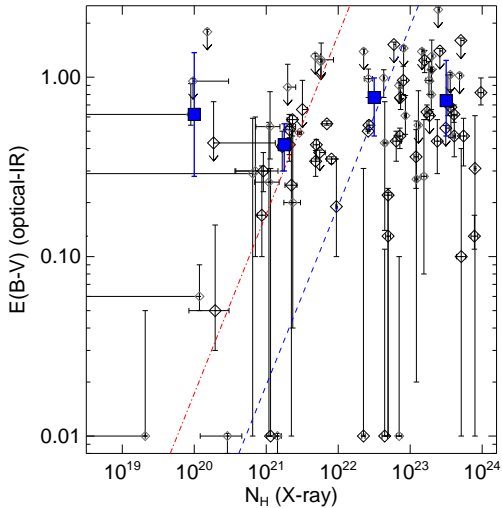}\label{dust_gas_ratio}
%\plotone{ebv_nh_pg1.ps}\label{dust_gas_ratio}
%\plotone{ebv_nh.ps}\label{dust_gas_ratio}
\caption{\ebv\ derived from dust reddened model SED fits
vs. X-ray-determined gas column density. Larger diamonds indicate
\ebv\ calculated using at least 3 ACS filters; smaller points indicate
that the \ebv\ calculation used only one or 2 filters. Average values
for four $N_H$ bins (\emph{blue squares}) show no strong trend:
intrinsically obscured AGN ($N_H > 10^{22}~\colden$) tend towards
larger reddening coefficients, but not universally.  The ratios span a
wide range, from approximately that of the SMC (\emph{red dot-dash},
$\mebv/N_H = 2.2\times 10^{-23}$~cm$^2$/H) to ratios in excess of
Galactic (\emph{blue dashed}, $\mebv/N_H = 1.7\times
10^{-22}$~cm$^2$/H).  }
\end{figure}

Our point-source fitting method produces estimates of \ebv\ from
optical point-source dust reddening. We also have gas column density
($N_H$) measurements from X-ray spectral slope fitting. These two
quantities allow for an independent measurement of the intrinsic
dust-to-gas ratio in AGN.  Figure~\ref{dust_gas_ratio} shows the
measures of dust vs. gas in the form of \ebv\ vs. $N_H$.  Binning the
data in Figure~\ref{dust_gas_ratio} over four $N_H$ ranges shows no
strong trend; however, a Kolmogorov-Smirov (K-S) test indicates only a
0.4\% chance that the subsample of \ebv\ values for objects with $N_H
< 10^{22}~{\rm cm}^{-2}$ is drawn from the same parent sample as the
subsample of \ebv\ values for objects with $N_H > 10^{22}~{\rm
cm}^{-2}$. This indicates that dust and gas obscuration may be weakly
correlated, so that sources with higher $N_H$ tend toward higher \ebv\
values, but with large scatter. The dust-to-gas ratios span a
relatively wide range, including dust-to-gas ratios similar to the
Galactic value, $\mebv /N_H = 1.7 \times 10^{-22}\mbox{
cm}^2/\mbox{H}$ \citep[see][]{draine03,shullvs85}, and those closer to
that in the SMC, $\mebv /N_H = 2.2 \times 10^{-23}\mbox{
cm}^2/\mbox{H}$ \citep{martin89}.

The SMC-like dust ratios are consistent with observations of local
Seyferts, based on their star formation histories and intrinsic
metallicities \citep{maiolino01a,willott04,hopkins04_AJ128}.  However,
as discussed in \citet{willott04}, the similarity of these dust-to-gas
ratios is likely a coincidence, due to the drastically different
physical conditions between AGN and the SMC.  \citet{maiolino01a,
maiolino97} discuss at least three causes for the much higher
gas-to-dust ratio relative to the Milky Way: (1) a difference in dust
grain composition and size such that UV absorption is less effective;
(2) the ratio of dust to gas is simply lower than in our galaxy; and
(3) an effect of the majority of X-ray absorption occurring very close
to the ionizing radiation and within the dust sublimation radius
(defined at roughly 1500 K, and on the order of
$R_{sub}=0.1-1$~pc). 
The work of \citet{elitzur06} supports this last explanation.

The majority (90\%) of our sample lack detected broad emission lines,
which within the AGN unification paradigm is commonly interpreted as
the result of an optically thick line-of-sight to the broad-line
region \citep[also within the dust sublimation radius;][]{honig06}.
This, combined with the high obscuration for much of our sample,
suggests most of the gas and dust along the line of sight lies within
the X-ray dissociation region.

\section{Eddington Ratios}\label{ledd}

\begin{figure}
\figurenum{9}
%\epsscale{1.0}
\plotone{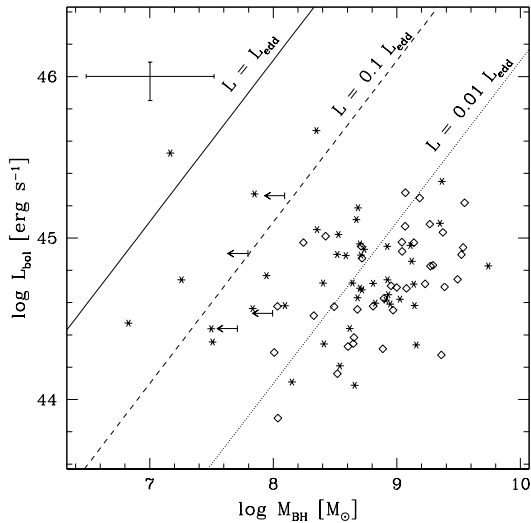}\label{mass_lum}
%\plotone{Mbh_lbol.eps}\label{mass_lum}
\caption{Bolometric AGN luminosity vs. black hole mass.
Bulge-dominated (\emph{diamonds}) and disk-dominated systems
(\emph{stars}) have overlapping distributions, although most of the
lowest-mass black holes are in disk-dominated systems.  Typical errors
are indicated at the upper left. Lines show 100\%, 10\%, and 1\% of
the Eddington ratio; the mean and median Eddington ratios for our
sample are $L = 0.009\times L_{Edd}$ and $L = 0.006\times L_{Edd}$,
respectively.  }
\end{figure}

The use of hard X-ray luminosity to select a complete sample of AGN is
vital to studies of AGN accretion properties.  Approximately 90\% of
our sources would fail to be included in samples selected by optical
spectroscopy.  Additionally, the use of \emph{HST} data allows for
black hole mass estimates that are independent of AGN properties. Most
AGN host galaxies are unresolved in black hole surveys selected by
broad $H\beta$ \citep{kollmeier06,vestergaard06,bentz06}, [O~III]
\citep{netzer07b} or CIV emission lines
\citep{vestergaard02,kaspi05}. Conversely, most of the optical spectra
in narrow-line AGN \citep{barger03,cowie03,szokoly04} are good enough
for redshifts, but lack the signal-to-noise for a measurement of \mbh\
\citep[\emph{e.g.},][]{heckman04}. Our X-ray selection criteria,
combined with not being dependent on line widths for \mbh\ estimates,
allows us to examine AGN that are not individually contributing
strongly to the accretion history of AGN, but that are a significant
fraction of the X-ray luminosity function (and hence total black hole
accretion) at intermediate redshifts.

Figure~\ref{mass_lum} shows the comparison between the black hole mass
(Section~\ref{bhmass}) and bolometric luminosity
(Section~\ref{dust_lum}).  The mean and median Eddington ratios for
our sample are $\left<\mlbol/L_{Edd}\right> = 0.009$ and $0.006$,
respectively (where $L_{Edd} \equiv 1.3\times10^{38}(\mmbh
/M_\odot)~\lum$), meaning the majority of the sample consists of AGN
with highly sub-Eddington accretion rates. Only one-third of the
sample has $\mlbol/L_{Edd} > 0.01$. This result -- a typically low
Eddington ratio within a sample that spans a wide range of values --
is similar to results for local Seyfert galaxies \citep[\emph{e.g.},
][]{cardamone07, mushotzky08}.

The scarcity of AGN with high accretion rates is at least partly to be
expected from our selection criteria.  By excluding AGN with low
host-to-point source luminosity ratios, we preferentially exclude
high-Eddington-ratio black holes for all but the smallest SMBH. Our
selection process (\S \ref{selection}) rejects 34 AGN+hosts in total,
of which 16 are too faint to accurately recover bulge
luminosities. Given their low fluxes, those 16 objects are unlikely to
be bright, near-Eddington accreting AGN. The other 18 rejected objects
are point-like in the optical images, which could possibly be due to
bright (near-Eddington-accreting) nuclei. However, even if those
objects were included, they would constitute a minority of the total
sample, indicating that our sample of sub-Eddington AGN represents the
dominant AGN population within the GOODS fields at $z < 1.25$.

The result that more than 90\% of our sample is accreting at less than
10\% of the Eddington limit is robust to possible sources of error in
our analysis. If we were to assume that our sources are accreting at
the Eddington limit and use that assumption to calculate SMBH masses,
our masses would decrease by more than two orders of magnitude, on
average. Such a large deviation is inconsistent both with independent
observations of black hole masses and bulges from $0 \lesssim z
\lesssim 3$ \citep[\emph{e.g.},][]{woo08,jahnke09,merloni10} and with
constraints on the maximum evolution of the black hole-bulge relation
from studies that \emph{do} assume Eddington-limited accretion
\citep{borys05,bluck11}.  Our results are also inconsistent with
scenarios in which most SMBHs cycle between a fully quiescent state
and a near-Eddington accretion state
\citep[\emph{e.g.},][]{king10}. Even assuming the largest possible
deviation in black hole masses based on the maximum evolution of the
bulge-black hole relation, our AGN are still in a phase of very slow
growth.

\begin{figure*}
\figurenum{10}
\epsscale{0.9}
\plotone{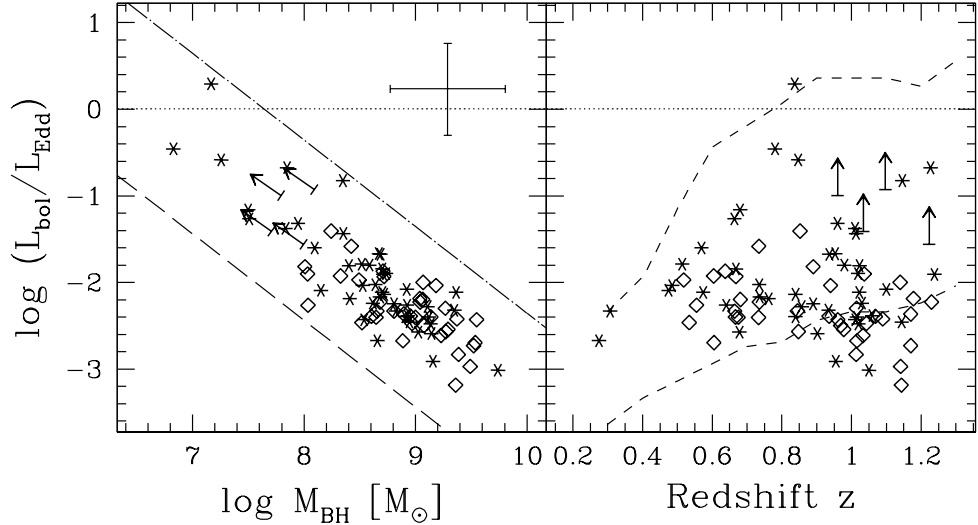}\label{edd_ratio}
%\plotone{l_bol_emp_bc_edd_ratios_adds.ps}\label{edd_ratio} - NOT used
%\plotone{Ledd_MBH_z.eps}\label{edd_ratio}
\caption{Eddington ratio vs. black hole mass (\emph{left}) and
redshift (\emph{right}). Symbols indicate bulge-like (\emph{diamonds})
and disk-like (\emph{stars}) host galaxy morphologies.  The dashed
line in the left panel indicates the lower $L_X$ selection criteria of
$3\times10^{42}~\lum$, and the dot-dashed line indicates the upper
luminosity limit for a single-source detection based on the
\citet{croom04} quasar luminosity function calculated for the GOODS
area. The dashed lines on the right indicate the maximum and minimum
luminosities of X-ray sources detected in the GOODS parent sample at
each redshift, assuming $\log \mmbh = 7.0 M_\odot$. In both panels,
the observations more or less fill the allowed area. }
\vspace{0.2in}
\end{figure*}

The Eddington ratios are presented in Figure~\ref{edd_ratio} in
relation to \mbh\ and redshift. To determine whether we see any
significant evolution of Eddington ratios \citep[as others have;
\emph{e.g.,}][]{netzer07c,gh07b}, we consider the relevant selection
biases. In Figure~\ref{edd_ratio}, we plot the hard X-ray luminosity
limit ($L_X > 3\times10^{42}~\lum$, dashed line) of our sample, which
excludes low Eddington ratio, low mass AGN. At the same time,
high-luminosity AGN are sufficiently rare that they do not appear in
pencil-beam surveys such as GOODS; we show this by plotting in
Figure~\ref{edd_ratio} the luminosity at which we expect to detect
only 1 AGN within the GOODS volume (which is a function of redshift)
from the quasar luminosity function \citep{croom04}.  After accounting
for these selection effects, we detect no correlation between
Eddington ratio and either black hole mass or redshift. This lack of
real trend is consistent with \citet{woo02}, who found that selection
effects can create the appearance of correlations where none
intrinsically exist.

Our values of $\mlbol/L_{Edd}$ are somewhat lower than those reported
by others \citep{kollmeier06,gh07b,netzer07}. Direct comparison
between different studies is complicated, however, due to the varying
selection criteria, redshifts and flux limits of each study. For
example, restricting a comparison to the objects in their paper with
redshifts similar to ours, \citet{kollmeier06} find that the
broad-line AGN from the AGES survey \citep{kochanek04} radiate at
near-Eddington rates, $0.1< \lambda < 1.0$.  But their study has
considerably brighter flux limits than our own, with AGES being 50\%
complete at $R=21.5$, whereas our sample is nearly complete down to
$z_{850}=24.0$ under the constraints outlined in
Sections~\ref{selection} and \ref{galfit}. Direct comparison to the
large, optically-selected SDSS quasar sample \citep[][]{schneider07}
is also difficult owing to that sample being highly incomplete at low
Eddington ratios \citep{kelly10}. Samples with similar selection
criteria and flux limits \citep{bundy08}, and those with overlapping
samples from the same survey \citep{ballo07}, report Eddington ratios
more similar to ours.

Interestingly, many of the aforementioned black hole samples at
different redshifts and with different selection criteria do not
appear to have radically different accretion rates. \citet{gh07b} show
that a sample of broad-line AGN selected from the SDSS have a peak
black hole mass of $10^7 M_\odot$ at $z<0.3$, with a typical Eddington
rate of 10\%, or a typical accretion rate of $2 \times
10^{-3}/\epsilon~M_{\odot}$~yr$^{-1}$, where $\epsilon$ is the
efficiency of converting mass to light.  Most AGN in our sample have
similar accretion rates (to within 40\%), despite having higher masses
and lower Eddington rates. Our accretion rates are consistent with
those of \citet{ballo07} sample, which overlaps ours somewhat, even
though they calculate \mbh\ and \lbol\ differently. Type 1 AGN from
the zCOSMOS sample \citep{merloni10}, which probes higher redshifts
and luminosities than our GOODS sample, are accreting at rate higher
than ours by only a factor of $\sim 4$. And \citet{hickox09}, who
probe a slightly lower-mass part of the black-hole mass function at
similar redshifts, find that black holes in AGN are accreting at a
rate within a factor of 2 of ours. These different samples of AGN span
a wide range of redshifts, luminosities, and black hole masses, yet
they have similar estimates of accretion rates in
$M_{\odot}$~yr$^{-1}$, assuming the same radiative efficiency. This
suggests that the accretion rates of AGN may be more directly related
to the supply of material from the circumnuclear region than the
properties of the central black hole.

If we follow \citet{elvis02} and assume a minimum radiative efficiency
of $\epsilon = 0.15$, our median black hole with an Eddington ratio of
0.006 and mass of $\MBHmed$ is accreting at $\dot{M} \sim
0.05~M_{\odot}~\rm yr^{-1}$. This accretion rate is too low to have
been the typical rate for the duration of the AGN activity cycle:
given our median redshift, it would take a seed black hole a minimum
of $\sim 150$\% of the age of the universe to that point to have grown
to its present size, assuming growth from the largest initial mass of
direct-collapse black holes \citep{volonteri08}, $\sim
10^{6}~M_{\odot}$. Given that the seed mass is likely to have been
lower, and that this accretion rate is super-Eddington for $\mmbh
\lesssim 3 \times 10^6~M_{\odot}$, the actual time would likely be far
higher. For any reasonable assumptions, therefore, the accretion rate
of a typical black hole in our sample must have been significantly
higher at some point in the past in order for it to grow to its
present size within a Hubble time. Assuming a larger radiative
efficiency, which Elvis et al. suggest is possible, only strengthens
this conclusion.

This is unsurprising given that black hole growth scenarios involving
a period of rapid (Eddington-limited) growth followed by a
power-law-decay growth rate \citep[\emph{e.g.},][]{hopkins06,yu08} are
more consistent with observations of Eddington ratio distributions in
a complete sample than constant-growth or ``light bulb'' scenarios
\citep{hopkins09b}. Under the specific timescales for a
self-regulating feedback model described in \citet{hopkins09b}, we
calculate that a black hole with our median observed redshift ($z =
0.94$) and Eddington ratio has an accretion rate that peaks at $z
\approx 1.1$, assuming a single episode of growth. Such an object
would easily be detected above our flux limits. The progenitor to our
median AGN should therefore be detected within our sample at its peak
growth stage (\emph{i.e.}, accreting at a high Eddington
ratio). Extending these calculations to each individual object in our
sample, we should detect $\approx 60$ progenitor AGN accreting at the
Eddington limit (within our limiting redshift, $z < 1.25$). 

In fact, we detect only 7 possible candidates: 3 sources consistent
with $\mlbol/L_{Edd}=1$ within our $1 \sigma$ uncertainties and 4
sources for which the detected $\mlbol/L_{Edd}$ is a lower limit. The
number of Eddington-limited AGN in our sample is too low by at least
an order of magnitude to be consistent with the predictions of the
\citeauthor{hopkins09b} model. We also considered the possibility that
the progenitors of the slow-growing AGN within our limiting redshift
($z < 1.25$) were removed from our sample due to their bright nuclei
rendering them point-like in the optical images (\S
\ref{selection}). However, not only are there not enough of those
objects (18) to account for the predictions of the model, but the
redshift distribution of the excluded point-like AGN is actually
\emph{lower} than that of the objects within our sample (median
$z=0.84$ vs. 0.94 in the included sample). If these objects were
progenitors to our slow-growing AGN, their redshifts would necessarily
be higher. Comparing the redshift distribution of expected progenitors
from our Monte Carlo simulations to the sample of excluded AGN, we
find the same result. We therefore argue that the timescales of the
single-episode model in \citet{hopkins09b} do not describe the
accretion history of our AGN.  A significant fraction of our observed
slow-growing black holes must have had a peak growth epoch beyond our
maximum redshift.

Overall, most of the X-ray selected AGN constituting our sample are
best described as being AGN with high black hole masses and with low
accretion rates, in a slow-growth phase. In order to grow to the sizes
we observe, the typical source in our sample must have had a
significantly higher accretion rate at some point well before $z \sim
1$. Although a preponderance of slowly growing AGN is qualitatively
consistent with self-regulating feedback models in which AGN spend
most of their time at low accretion rates, our result is inconsistent
with quantitative timescales predicted by common parameterizations of
this model. The growth history of black holes is clearly more complex
than that described by a single simple model, a conclusion echoed by
other studies examining AGN at a variety of different redshifts and
luminosities \citep{kauffmann09,merloni10,cardamone10,steinhardt11}.

\section{Conclusions}

We studied 87 AGN with $z < 1.25$ from the GOODS survey to understand
the fundamental properties of moderate-luminosity obscured AGN that
make up most of the X-ray background and thus a large fraction of
black hole growth.  Most of these AGN would not be selected via broad
emission line surveys or blue color excesses, \emph{i.e.}, the
highest-$L$ and $L/L_{edd}$ objects are excluded. However, this sample
represents the bulk of AGN emission at $z \lesssim 1.25$.

After morphological fits to separate AGN/host galaxy light and
determine bulge-to-total ratios, we estimated black hole masses for
each AGN using the $\mmbh-L_{B}$ relation of \citet{marconi_hunt03} as
reformulated by \citet{ff05}. We do not assume this relation evolves
with redshift, but use a Monte Carlo error analysis to account for
uncertainty in the time evolution of Equation~\ref{ML_equation} as
well as other parameter uncertainties. The resulting uncertainties in
our black hole masses are typically on the order of 0.5~dex.

Using an unobscured AGN template calculated from broad-line AGN in the
GOODS fields, we calculated bolometric luminosities individually for
each object by reddening the normalized template SED to match nuclear
point-source colors and using the fitted \ebv\ to calculate the energy
re-radiated in the far-infrared by dust surrounding the AGN. This
method is consistent with single-prescription empirical bolometric
correction methods and with independent theoretical models, but has
less scatter.

The SED fitting of the nuclear point sources gives the added benefit
of an estimate of dust absorption upon the AGN template.  Assuming a
standard Galactic reddening curve, we find that 70\% of the AGN have
dust-to-gas ratios similar to
%considerably less than Galactic, and even less than
%the SMC, like 
local Seyferts \citep{maiolino01a}.  The host galaxy contribution to
the nuclear line-of-sight obscuration in the GOODS AGN is minimal for
most of our sources.

Our survey reveals a considerable fraction of low-Eddington-accreting,
high-black-hole-mass AGN in normal host galaxies. The mean Eddington
ratio for our sample is $L/L_{Edd} \sim 0.009_{-0.005}^{+0.046}$,
considerably lower than in broad-line AGN surveys such as SDSS
\citep{netzer07b,gh07b}.  Many of these AGN have properties consistent
with local Seyfert AGN \citep{mushotzky08}. Even considering the small
uncertainties in the Eddington ratio, which reflect the current
uncertainty in the time evolution of the bulge-black hole mass
relation, our typical source is still accreting at well below the
Eddington limit.

The observed Eddington ratios of our sources are low, but their black
hole masses are high, meaning they must have been accreting at
significantly higher rates at some point during their growth history
in order for them to have reached their observed masses within a
Hubble time. According to self-regulating feedback accretion models
for black hole growth, our black holes could be the slow-growth phase
of their AGN lifetimes; comparison with accretion timescales from
simple models of self-regulated feedback indicate that our AGN sample
may have a complex accretion history. A significant fraction of our
sample must have been accreting at near-Eddington rates before $z \sim
1$. Such a complex growth history is consistent with the downsizing
scenario \citep{barger05,hasinger05}.

The use of the point-source luminosity as leverage to extract the
multi-wavelength AGN SED from a combined AGN plus host galaxy source
is a promising technique. In particular, the high-resolution infrared
data promised by \emph{JWST} could be used to significantly increase
the accuracy of the central reddening determination, which would
decrease uncertainty about the re-radiated dust luminosity in the FIR
range. This technique may also benefit from the separation of AGN and
host galaxies at IR wavelengths using WFC3
\citep{schawinski11}. Although the efficacy of host-point-source
separation has not yet been studied in as much depth using WFC3 data
as it has for ACS, the addition of several data points to the
low-resolution point-source SED would be very valuable.

\acknowledgments

The authors wish to thank to C. Peng for making {\tt GALFIT} publicly
available, and for many enlightening discussions.  Thanks also to
%E. Treister for sharing his model-generated bolometric luminosity
%tables, and to 
C. Cardamone and E. Bonning for helpful comments on this
manuscript. We appreciate helpful comments from the anonymous
referee. The JavaScript Cosmology Calculator was used while preparing
this paper \citep{wright06}. The authors acknowledge support from NASA
through grants HST-AR-10689.01-A, HST-GO-09425.13-A and
HST-GO-09822.09-A from the Space Telescope Science Institute, which is
operated by the Association of Universities for Research in Astronomy
under NASA contract NAS 5-26555.  The authors also acknowledge support
from the \emph{XMM-Newton} Grant NNG06GD72G. \emph{XMM-Newton} is an
ESA science mission with instruments and contributions directly funded
by ESA Member States and the USA (NASA). Support for the work of
E.T. was provided by NASA through Chandra Postdoctoral Fellowship
grant number PF8-90055, issued by the Chandra X-ray Observatory
Center, which is operated by the Smithsonian Astrophysical Observatory
for and on behalf of NASA under contract NAS8-03060.

\bibliographystyle{apj}
%\bibliography{refs}
\bibliography{apj-jour,bhmass_lbol,refs}

%%%%%%%%%%%%%%
% Tables
%%%%%%%%%%%%%%

%  Source listing
\pagebreak
\LongTables

%%%%%%%%%%%%%%%%%%%%%%%%%%%% Table 1 %%%%%%%%%%%%%%%%%%%%%%%%%%%%%%

\begin{deluxetable}{lcccccccc}
\label{bhm_table}
%\begin{minipage}{7.0in}
\tablehead{%\multicolumn{9}{c}{Source List and Black Hole Mass estimates} \\
   &  \multicolumn{2}{c}{ Optical Position} & \multicolumn{1}{c}{}  & \multicolumn{3}{c}{ Rest-frame B (AB mag)}  &  &  \\
\colhead{  ID\tablenotemark{a}} &\colhead{ RA} &\colhead{  DEC} &\colhead{  $z$} &\colhead{  Total} &\colhead{  Pt src} &\colhead{  Host} & \colhead{  B/Tot} &\colhead{  $\log M_{BH}$ ($\rm{M}_{\odot}$)} 
%\hline
}
\tablecaption{GOODS AGN sample with point source and host galaxy deconvolution for objects between $0.2<z<1.25$.}

\tablecolumns{9}

\startdata 
N 48 & 188.983765 & 62.205441 & 0.940 & 21.741$\pm$0.012 & \nodata & 21.74$\pm$0.01 &	$0.79_{-0.30}^{+0.17}$ &	$9.18_{-0.51}^{+0.52}$ \\
N 72 & 189.023956 & 62.167603 & 0.936 & 22.020$\pm$0.007 & 23.40$\pm$0.19 & 22.38$\pm$0.23 &	$0.48_{-0.26}^{+0.28}$ &	$8.69_{-0.50}^{+0.51}$ \\
N 76 & 189.027634 & 62.164326 & 0.637 & 21.389$\pm$0.010 & 26.79$\pm$0.39 & 21.40$\pm$0.21 &	$0.53_{-0.27}^{+0.27}$ &	$8.71_{-0.50}^{+0.50}$ \\
N 82 & 189.033875 & 62.176731 & 0.681 & 20.990$\pm$0.004 & 22.90$\pm$0.24 & 21.19$\pm$0.24 &	$0.82_{-0.30}^{+0.15}$ &	$9.07_{-0.52}^{+0.52}$ \\
N 90 & 189.046814 & 62.150940 & 1.140 & 22.340$\pm$0.013 & 24.48$\pm$0.14 & 22.50$\pm$0.20 &	$0.77_{-0.29}^{+0.18}$ &	$9.07_{-0.52}^{+0.52}$ \\
N 93 & 189.050156 & 62.194221 & 0.275 & 19.784$\pm$0.001 & 23.08$\pm$0.21 & 19.84$\pm$0.23 &	$0.83_{-0.30}^{+0.14}$ &	$8.66_{-0.50}^{+0.50}$ \\
N103 & 189.060364 & 62.121876 & 0.969 & 21.194$\pm$0.012 & 24.86$\pm$0.15 & 21.30$\pm$0.18 &	$0.39_{-0.23}^{+0.29}$ &	$9.08_{-0.52}^{+0.52}$ \\
N110 & 189.070221 & 62.104027 & 1.141 & 21.479$\pm$0.008 & 26.23$\pm$0.26 & 21.48$\pm$0.19 &	$0.76_{-0.29}^{+0.19}$ &	$9.49_{-0.55}^{+0.55}$ \\
N113 & 189.071228 & 62.169903 & 0.845 & 22.451$\pm$0.011 & 25.24$\pm$0.17 & 22.45$\pm$0.17 &	$0.87_{-0.30}^{+0.12}$ &	$8.81_{-0.51}^{+0.50}$ \\
N116\tablenotemark{b} & 189.077530 & 62.187622 & 1.022 & 20.554$\pm$0.006 & 22.53$\pm$0.29 & 20.55$\pm$0.26 &	$0.32_{-0.21}^{+0.33}$ &	$9.36_{-0.54}^{+0.54}$ \\
N127 & 189.088516 & 62.185921 & 1.014 & 21.881$\pm$0.011 & 26.22$\pm$0.26 & 21.88$\pm$0.19 &	$0.89_{-0.30}^{+0.10}$ &	$9.27_{-0.53}^{+0.53}$ \\
N139 & 189.098541 & 62.310368 & 1.013 & 21.411$\pm$0.014 & 24.67$\pm$0.14 & 21.41$\pm$0.19 &	$0.75_{-0.30}^{+0.19}$ &	$9.39_{-0.54}^{+0.54}$ \\
N150 & 189.114609 & 62.174030 & 0.762 & 23.115$\pm$0.023 & 26.69$\pm$0.38 & 23.11$\pm$0.19 &	$0.86_{-0.30}^{+0.12}$ &	$8.41_{-0.49}^{+0.50}$ \\
N158 & 189.121506 & 62.179565 & 1.013 & 21.957$\pm$0.018 & 27.88$\pm$0.76 & 21.96$\pm$0.17 &	$0.13_{-0.07}^{+0.30}$ &	$8.35_{-0.50}^{+0.49}$ \\
N160 & 189.122055 & 62.270576 & 0.848 & 21.464$\pm$0.005 & 24.00$\pm$0.14 & 21.46$\pm$0.22 &	$0.95_{-0.30}^{+0.05}$ &	$9.27_{-0.53}^{+0.53}$ \\
N163 & 189.124893 & 62.095203 & 0.485 & 21.389$\pm$0.005 & 23.81$\pm$0.16 & 21.39$\pm$0.22 &	$0.69_{-0.29}^{+0.22}$ &	$8.53_{-0.50}^{+0.50}$ \\
N164 & 189.125244 & 62.156734 & 0.953 & 22.425$\pm$0.009 & 25.23$\pm$0.16 & 22.42$\pm$0.17 &	$0.46_{-0.26}^{+0.28}$ &	$8.67_{-0.50}^{+0.50}$ \\
N170 & 189.132751 & 62.295914 & 0.680 & 21.829$\pm$0.008 & 26.15$\pm$0.24 & 21.85$\pm$0.19 &	$0.05_{-0.03}^{+0.22}$ &	$7.50_{-0.53}^{+0.52}$ \\
N174 & 189.138474 & 62.143036 & 0.934 & 21.084$\pm$0.005 & 22.52$\pm$0.30 & 21.42$\pm$0.26 &	$0.86_{-0.30}^{+0.12}$ &	$9.35_{-0.54}^{+0.54}$ \\
N177 & 189.140244 & 62.168388 & 1.016 & 21.336$\pm$0.008 & 26.40$\pm$0.30 & 21.35$\pm$0.19 &	$0.39_{-0.24}^{+0.29}$ &	$9.12_{-0.52}^{+0.52}$ \\
N187 & 189.145264 & 62.274620 & 0.847 & 22.444$\pm$0.015 & 25.77$\pm$0.19 & 22.50$\pm$0.18 &	$0.03_{-0.03}^{+0.34}$ &	$7.26_{-0.54}^{+0.54}$ \\
N194 & 189.153198 & 62.199001 & 0.555 & 22.969$\pm$0.008 & 27.24$\pm$0.52 & 22.99$\pm$0.19 &	$0.73_{-0.29}^{+0.20}$ &	$8.04_{-0.50}^{+0.50}$ \\
N201 & 189.160385 & 62.227596 & 1.020 & 22.059$\pm$0.015 & 25.84$\pm$0.19 & 22.09$\pm$0.18 &	$0.81_{-0.30}^{+0.16}$ &	$9.14_{-0.52}^{+0.52}$ \\
N205 & 189.162918 & 62.162346 & 1.230 & 22.998$\pm$0.017 & 26.16$\pm$0.27 & 23.06$\pm$0.19 &	$1.00_{-0.30}^{+0.00}$ &	$9.04_{-0.51}^{+0.51}$ \\
N217 & 189.173203 & 62.163483 & 0.518 & 21.497$\pm$0.005 & 28.74$\pm$1.08 & 21.50$\pm$0.17 &	$0.60_{-0.35}^{+0.30}$ &	$8.49_{-0.50}^{+0.50}$ \\
N222 & 189.175827 & 62.262722 & 0.857 & 21.380$\pm$0.008 & 24.00$\pm$0.14 & 21.48$\pm$0.22 &	$0.66_{-0.32}^{+0.24}$ &	$9.11_{-0.52}^{+0.52}$ \\
N240 & 189.193115 & 62.234734 & 0.961 & 20.867$\pm$0.005 & 22.85$\pm$0.24 & 21.06$\pm$0.24 &	$0.89_{-0.30}^{+0.10}$ &	$9.55_{-0.55}^{+0.56}$ \\
N242 & 189.194107 & 62.149166 & 0.890 & 23.861$\pm$0.041 & 28.37$\pm$0.93 & 23.88$\pm$0.16 &	$0.49_{-0.28}^{+0.30}$ &	$8.01_{-0.50}^{+0.50}$ \\
N261 & 189.209030 & 62.204823 & 0.902 & 21.971$\pm$0.010 & 26.91$\pm$0.43 & 21.98$\pm$0.20 &	$1.00_{-0.30}^{+0.00}$ &	$9.14_{-0.52}^{+0.52}$ \\
N262 & 189.209595 & 62.334686 & 1.011 & 22.566$\pm$0.010 & 23.97$\pm$0.14 & 22.92$\pm$0.22 &	$0.47_{-0.26}^{+0.28}$ &	$8.54_{-0.50}^{+0.50}$ \\
N266 & 189.213638 & 62.181107 & 1.100 & 22.975$\pm$0.017 & 25.89$\pm$0.19 & 23.05$\pm$0.18 &	$1.00_{-0.30}^{+0.00}$ &	$8.92_{-0.51}^{+0.51}$ \\
N278 & 189.222992 & 62.338577 & 1.023 & 23.166$\pm$0.013 & 24.76$\pm$0.15 & 23.45$\pm$0.19 &	$0.56_{-0.28}^{+0.27}$ &	$8.40_{-0.50}^{+0.49}$ \\
N286 & 189.231064 & 62.219883 & 0.955 & 21.626$\pm$0.008 & 24.52$\pm$0.14 & 21.70$\pm$0.19 &	$0.69_{-0.29}^{+0.22}$ &	$9.16_{-0.52}^{+0.52}$ \\
N304 & 189.245163 & 62.243118 & 0.678 & 20.765$\pm$0.005 & 23.38$\pm$0.19 & 20.87$\pm$0.23 &	$0.56_{-0.28}^{+0.26}$ &	$9.02_{-0.51}^{+0.52}$ \\
N309 & 189.249268 & 62.326241 & 1.144 & 22.582$\pm$0.017 & 24.92$\pm$0.14 & 22.72$\pm$0.18 &	$0.71_{-0.30}^{+0.21}$ &	$8.95_{-0.51}^{+0.51}$ \\
N323\tablenotemark{b} & 189.261353 & 62.262199 & 0.514 & 20.393$\pm$0.002 & 21.72$\pm$0.41 & 20.77$\pm$0.54 &	$0.33_{-0.21}^{+0.31}$ &	$8.52_{-0.54}^{+0.54}$ \\
N349 & 189.281601 & 62.332314 & 1.030 & 23.522$\pm$0.027 & 26.52$\pm$0.33 & 23.59$\pm$0.19 &	$1.00_{-0.30}^{+0.00}$ &	$8.62_{-0.50}^{+0.50}$ \\
N352 & 189.282730 & 62.268341 & 0.936 & 20.980$\pm$0.006 & 26.69$\pm$0.37 & 20.99$\pm$0.20 &	$0.24_{-0.16}^{+0.30}$ &	$8.95_{-0.51}^{+0.51}$ \\
N370 & 189.300629 & 62.298428 & 1.060 & 22.683$\pm$0.014 & 24.89$\pm$0.14 & 22.84$\pm$0.18 &	$0.92_{-0.30}^{+0.08}$ &	$8.93_{-0.51}^{+0.51}$ \\
N373 & 189.307480 & 62.240242 & 0.475 & 22.628$\pm$0.006 & 28.17$\pm$0.85 & 22.63$\pm$0.16 &	$0.99_{-0.30}^{+0.00}$ &	$8.15_{-0.50}^{+0.50}$ \\
N384 & 189.316345 & 62.203796 & 1.019 & 21.985$\pm$0.018 & 26.52$\pm$0.33 & 22.00$\pm$0.21 &	$0.31_{-0.20}^{+0.30}$ &	$8.74_{-0.50}^{+0.50}$ \\
N390 & 189.319489 & 62.292667 & 1.146 & 22.123$\pm$0.008 & 22.98$\pm$0.22 & 22.78$\pm$0.23 &	$0.20_{-0.13}^{+0.32}$ &	$8.35_{-0.50}^{+0.50}$ \\
N402 & 189.334732 & 62.231556 & 0.780 & 23.014$\pm$0.232 & 29.79$\pm$0.85 & 23.02$\pm$0.05 &	$0.02_{-0.02}^{+0.20}$ &	$6.83_{-0.58}^{+0.58}$ \\
N405 & 189.340988 & 62.176670 & 0.978 & 21.181$\pm$0.006 & 25.00$\pm$0.14 & 21.18$\pm$0.17 &	$0.55_{-0.27}^{+0.26}$ &	$9.29_{-0.53}^{+0.53}$ \\
N437 & 189.393890 & 62.232430 & 0.839 & 20.890$\pm$0.006 & 25.23$\pm$0.15 & 20.91$\pm$0.17 &	$0.27_{-0.17}^{+0.30}$ &	$8.91_{-0.51}^{+0.51}$ \\
N442 & 189.402100 & 62.225620 & 0.852 & 22.193$\pm$0.009 & 24.76$\pm$0.14 & 22.30$\pm$0.18 &	$0.21_{-0.14}^{+0.30}$ &	$8.24_{-0.50}^{+0.50}$ \\
N448 & 189.408051 & 62.219257 & 1.238 & 22.709$\pm$0.016 & 26.36$\pm$0.29 & 22.75$\pm$0.20 &	$0.36_{-0.22}^{+0.29}$ &	$8.71_{-0.50}^{+0.50}$ \\
N451\tablenotemark{b} & 189.413559 & 62.349976 & 0.837 & 20.528$\pm$0.006 & 21.58$\pm$0.42 & 21.05$\pm$0.64 &	$0.01_{-0.01}^{+0.19}$ &	$7.17_{-0.63}^{+0.60}$ \\
N471 & 189.462021 & 62.266979 & 1.170 & 21.007$\pm$0.006 & 24.12$\pm$0.14 & 21.07$\pm$0.21 &	$0.52_{-0.27}^{+0.27}$ &	$9.52_{-0.55}^{+0.55}$ \\
N473\tablenotemark{b} & 189.469681 & 62.274593 & 0.307 & 19.467$\pm$0.001 & 20.67$\pm$0.47 & 19.90$\pm$1.12 &	$0.99_{-0.30}^{+0.01}$ &	$8.83_{-0.68}^{+0.71}$ \\
S 44 & 53.015217 & -27.767685 & 0.574 & 21.434$\pm$0.004 & 25.56$\pm$0.17 & 21.46$\pm$0.18 &	$0.71_{-0.29}^{+0.21}$ &	$8.70_{-0.50}^{+0.50}$ \\
S 84 & 53.050934 & -27.772406 & 1.033 & 21.866$\pm$0.010 & 25.50$\pm$0.17 & 21.91$\pm$0.18 &	$0.80_{-0.30}^{+0.16}$ &	$9.23_{-0.52}^{+0.53}$ \\
S 88 & 53.055191 & -27.711349 & 0.605 & 19.825$\pm$0.002 & 23.45$\pm$0.18 & 19.86$\pm$0.23 &	$0.90_{-0.30}^{+0.09}$ &	$9.54_{-0.55}^{+0.56}$ \\
S 91 & 53.057728 & -27.713583 & 0.735 & 21.240$\pm$0.006 & 25.68$\pm$0.18 & 21.26$\pm$0.18 &	$0.28_{-0.18}^{+0.30}$ &	$8.64_{-0.50}^{+0.50}$ \\
S103\tablenotemark{b} & 53.062420 & -27.857510 & 0.675 & 20.936$\pm$0.005 & 25.52$\pm$0.17 & 20.95$\pm$0.18 &	$0.27_{-0.17}^{+0.30}$ &	$8.65_{-0.50}^{+0.50}$ \\
S117 & 53.071434 & -27.717588 & 0.569 & 21.742$\pm$0.003 & 23.51$\pm$0.18 & 21.98$\pm$0.22 &	$0.31_{-0.20}^{+0.29}$ &	$8.09_{-0.50}^{+0.50}$ \\
S118 & 53.071533 & -27.872456 & 1.097 & 21.774$\pm$0.010 & 24.89$\pm$0.14 & 21.84$\pm$0.18 &	$<0.054$ &	$<8.09$ \\
S134 & 53.085327 & -27.792313 & 0.604 & 22.118$\pm$0.010 & 26.08$\pm$0.22 & 22.15$\pm$0.18 &	$0.53_{-0.27}^{+0.27}$ &	$8.33_{-0.50}^{+0.49}$ \\
S137 & 53.089264 & -27.708660 & 0.736 & 21.587$\pm$0.007 & 24.64$\pm$0.14 & 21.65$\pm$0.19 &	$0.44_{-0.25}^{+0.28}$ &	$8.68_{-0.50}^{+0.50}$ \\
S139 & 53.091618 & -27.782206 & 0.668 & 20.893$\pm$0.006 & 24.49$\pm$0.14 & 20.93$\pm$0.20 &	$0.30_{-0.19}^{+0.30}$ &	$8.70_{-0.50}^{+0.50}$ \\
S151 & 53.096489 & -27.765188 & 1.223 & 22.328$\pm$0.016 & 25.25$\pm$0.16 & 22.40$\pm$0.17 &	$<0.055$ &	$<8.00$ \\
S155 & 53.101063 & -27.690676 & 0.534 & 21.707$\pm$0.006 & 24.98$\pm$0.14 & 21.76$\pm$0.17 &	$0.76_{-0.30}^{+0.19}$ &	$8.52_{-0.50}^{+0.50}$ \\
S156 & 53.102264 & -27.669563 & 0.890 & 22.545$\pm$0.011 & 27.73$\pm$0.70 & 22.55$\pm$0.17 &	$0.84_{-0.30}^{+0.13}$ &	$8.81_{-0.50}^{+0.50}$ \\
S159 & 53.103516 & -27.933329 & 1.170 & 23.112$\pm$0.027 & 26.83$\pm$0.40 & 23.15$\pm$0.22 &	$0.89_{-0.30}^{+0.10}$ &	$8.90_{-0.51}^{+0.51}$ \\
S160 & 53.103989 & -27.835567 & 1.037 & 23.945$\pm$0.033 & 25.94$\pm$0.20 & 24.13$\pm$0.28 &	$0.45_{-0.27}^{+0.32}$ &	$8.03_{-0.51}^{+0.51}$ \\
S161 & 53.104088 & -27.683752 & 0.733 & 21.446$\pm$0.005 & 24.19$\pm$0.14 & 21.54$\pm$0.21 &	$0.81_{-0.30}^{+0.16}$ &	$9.00_{-0.51}^{+0.52}$ \\
S162 & 53.104607 & -27.845348 & 1.043 & 22.646$\pm$0.013 & 24.97$\pm$0.14 & 22.78$\pm$0.31 &	$0.99_{-0.30}^{+0.01}$ &	$8.97_{-0.52}^{+0.52}$ \\
S164 & 53.104836 & -27.913925 & 1.090 & 21.209$\pm$0.005 & 24.88$\pm$0.14 & 21.25$\pm$0.18 &	$0.52_{-0.27}^{+0.27}$ &	$9.37_{-0.54}^{+0.53}$ \\
S171 & 53.107746 & -27.918444 & 1.034 & 22.623$\pm$0.019 & 26.37$\pm$0.29 & 22.66$\pm$0.20 &	$<0.056$ &	$<7.71$ \\
S176 & 53.111511 & -27.695988 & 0.734 & 21.887$\pm$0.009 & 24.39$\pm$0.14 & 22.00$\pm$0.20 &	$0.35_{-0.22}^{+0.29}$ &	$8.42_{-0.49}^{+0.49}$ \\
S179\tablenotemark{b} & 53.115097 & -27.695805 & 0.668 & 21.614$\pm$0.005 & 24.83$\pm$0.14 & 21.67$\pm$0.18 &	$0.62_{-0.28}^{+0.25}$ &	$8.72_{-0.50}^{+0.50}$ \\
S184 & 53.120827 & -27.958441 & 0.640 & 20.284$\pm$0.004 & 21.78$\pm$0.40 & 20.60$\pm$0.48 &	$0.40_{-0.24}^{+0.29}$ &	$8.92_{-0.54}^{+0.55}$ \\
S193\tablenotemark{b} & 53.125252 & -27.756536 & 0.960 & 21.793$\pm$0.009 & 23.02$\pm$0.22 & 22.22$\pm$0.23 &	$<0.055$ &	$<7.79$ \\
S200 & 53.133675 & -27.698660 & 0.960 & 23.000$\pm$0.028 & 24.56$\pm$0.14 & 23.30$\pm$0.19 &	$0.21_{-0.14}^{+0.33}$ &	$7.95_{-0.51}^{+0.50}$ \\
S203 & 53.137436 & -27.688057 & 1.050 & 20.922$\pm$0.007 & 25.13$\pm$0.15 & 20.94$\pm$0.17 &	$0.98_{-0.30}^{+0.02}$ &	$9.74_{-0.57}^{+0.57}$ \\
S214 & 53.145634 & -27.919777 & 0.839 & 22.182$\pm$0.010 & 24.42$\pm$0.14 & 22.33$\pm$0.20 &	$0.65_{-0.29}^{+0.24}$ &	$8.72_{-0.50}^{+0.51}$ \\
S220 & 53.149342 & -27.683189 & 0.735 & 21.764$\pm$0.006 & 25.82$\pm$0.19 & 21.79$\pm$0.18 &	$0.50_{-0.27}^{+0.27}$ &	$8.68_{-0.50}^{+0.50}$ \\
S227 & 53.152973 & -27.735123 & 0.665 & 22.059$\pm$0.010 & 24.28$\pm$0.14 & 22.21$\pm$0.21 &	$0.89_{-0.30}^{+0.09}$ &	$8.65_{-0.50}^{+0.51}$ \\
S229 & 53.156075 & -27.666695 & 0.664 & 22.164$\pm$0.008 & 23.03$\pm$0.22 & 22.81$\pm$0.23 &	$0.13_{-0.07}^{+0.30}$ &	$7.51_{-0.53}^{+0.52}$ \\
S249 & 53.173805 & -27.724491 & 0.979 & 22.637$\pm$0.017 & 27.09$\pm$0.48 & 22.66$\pm$0.20 &	$0.45_{-0.26}^{+0.30}$ &	$8.59_{-0.50}^{+0.50}$ \\
S263\tablenotemark{b} & 53.185226 & -27.827835 & 1.016 & 22.308$\pm$0.008 & 24.25$\pm$0.14 & 22.51$\pm$0.21 &	$0.69_{-0.29}^{+0.22}$ &	$8.89_{-0.51}^{+0.51}$ \\
S271 & 53.195938 & -27.729589 & 1.178 & 22.146$\pm$0.022 & 25.17$\pm$0.15 & 22.22$\pm$0.18 &	$0.51_{-0.27}^{+0.27}$ &	$9.04_{-0.51}^{+0.52}$ \\
S273 & 53.196571 & -27.863205 & 1.069 & 22.252$\pm$0.009 & 24.70$\pm$0.14 & 22.37$\pm$0.19 &	$0.29_{-0.18}^{+0.29}$ &	$8.60_{-0.50}^{+0.50}$ \\
S276 & 53.200741 & -27.882389 & 0.667 & 20.303$\pm$0.004 & 23.04$\pm$0.22 & 20.39$\pm$0.23 &	$0.48_{-0.26}^{+0.28}$ &	$9.14_{-0.52}^{+0.53}$ \\
S286\tablenotemark{b} & 53.220360 & -27.855505 & 1.227 & 21.830$\pm$0.009 & 23.22$\pm$0.20 & 22.18$\pm$0.23 &	$0.03_{-0.03}^{+0.21}$ &	$7.85_{-0.51}^{+0.51}$ \\
S293 & 53.237385 & -27.835745 & 1.143 & 21.909$\pm$0.009 & 26.02$\pm$0.21 & 21.93$\pm$0.19 &	$0.86_{-0.30}^{+0.12}$ &	$9.36_{-0.53}^{+0.54}$ \\
S300 & 53.245888 & -27.861118 & 1.010 & 23.964$\pm$0.047 & 26.99$\pm$0.45 & 24.03$\pm$0.20 &	$0.28_{-0.19}^{+0.36}$ &	$7.83_{-0.51}^{+0.51}$ \\
\enddata
%\end{tabular}
%\end{minipage}
\tablenotetext{a}{X-ray IDs from \citet{a03}}
\tablenotetext{b}{Broad-Line AGN \citep{cowie03,wirth04,szokoly04}.}
\end{deluxetable}

%%%%%%%%%%%%%%%%%%%%%%%%%% End Table 1 %%%%%%%%%%%%%%%%%%%%%%%%%%%%

%%%%%%%%%%%%%%%%%%%%%%%%%%%% Table 2 %%%%%%%%%%%%%%%%%%%%%%%%%%%%%%

%\renewcommand{\thefootnote}{\alph{footnote}}
\label{lum_table}
\begin{deluxetable}{lccccccccc}
\tablecaption{Bolometric luminosities derived from reddened SEDs\tablenotemark{a}, $L_X$ correction\tablenotemark{b}, direct integration\tablenotemark{c}, and comparison model\tablenotemark{d}.}
\tablehead{ \colhead{ID} & \colhead{$\log L_X$\tablenotemark{e}} & \colhead{$\log N_H$\tablenotemark{f}} & \colhead{\ebv} & \colhead{$N_{pt}$} & \colhead{\llbol\tablenotemark{a, e}} & \colhead{\llbolX\tablenotemark{b, e}} & \colhead{\llsum\tablenotemark{c, e}} & \colhead{\lltuv\tablenotemark{d, e}} & \colhead{$\log \lambda_{edd}$\tablenotemark{g}} 
%\\& \tiny [erg s$^{-1}$] & \tiny [cm$^{-2}$] & & &\tiny [erg s$^{-1}$] &\tiny [erg s$^{-1}$] &\tiny [erg s$^{-1}$] &\tiny [erg s$^{-1}$] & 
}
\tablecolumns{10}
\startdata
N 48 & 44.08 & 23.30 & 1.10$^{+ 0.04}_{- 0.15}$ & 2 & 45.25 & 45.24 & 45.25 &	45.90 & 	$-2.04_{-0.52}^{+0.52}$ \\
N 72 & 42.83 & 22.85 & 0.01$^{+ 0.09}_{- 0.00}$ & 2 & 45.19 & 43.99 & 45.02 &	44.40 & 	$-1.68_{-0.55}^{+0.54}$ \\
N 76 & 44.21 & 23.98 & 0.82$^{+ 0.17}_{- 0.12}$ & 3 & 44.95 & 45.36 & 45.05 &	46.05 & 	$-1.87_{-0.50}^{+0.50}$ \\
N 82 & 42.61 & 22.69 & 0.13$^{+ 0.11}_{- 0.12}$ & 3 & 45.07 & 43.77 & 45.25 &	44.14 & 	$-2.20_{-0.56}^{+0.54}$ \\
N 90 & 43.54 & 23.09 & 0.27$^{+ 0.30}_{- 0.25}$ & 1 & 45.28 & 44.70 & 45.34 &	45.25 & 	$-2.00_{-0.56}^{+0.54}$ \\
N 93 & 42.60 & 21.32 & 0.42$^{+ 0.04}_{- 0.05}$ & 4 & 44.09 & 43.76 & 44.72 &	44.13 & 	$-2.67_{-0.50}^{+0.50}$ \\
N103 & 42.54 & 20.27 & 0.43$^{+ 0.30}_{- 0.42}$ & 3 & 44.69 & 43.70 & 45.26 &   	44.06 & 	$-2.50_{-0.52}^{+0.52}$ \\
N110 & 42.77 & 20.81 & 0.29$^{+ 0.30}_{- 0.28}$ & 2 & 44.74 & 43.93 & 45.32 &   	44.33 & 	$-2.97_{-0.59}^{+0.57}$ \\
N113 & 43.15 & 21.32 & 0.51$^{+ 0.11}_{- 0.17}$ & 3 & 44.58 & 44.31 & 45.02 &   	44.79 & 	$-2.33_{-0.50}^{+0.51}$ \\
N116\tablenotemark{h} & 43.92 & 20.29 & 0.05$^{+ 0.10}_{- 0.02}$ & 4 & 45.35 & 45.07 & 45.60 &   	45.70 & 	$-2.11_{-0.55}^{+0.55}$ \\
N127 & 43.52 & 23.52 & 0.74$^{+ 0.29}_{- 0.13}$ & 2 & 45.09 & 44.68 & 45.24 &   	45.23 & 	$-2.30_{-0.53}^{+0.53}$ \\
N139 & 43.00 & 21.04 & 0.26$^{+ 0.30}_{- 0.25}$ & 2 & 44.70 & 44.16 & 45.25 &   	44.60 & 	$-2.83_{-0.54}^{+0.54}$ \\
N150 & 42.97 & 22.93 & 0.61$^{+ 0.18}_{- 0.13}$ & 2 & 44.34 & 44.12 & 44.40 &   	44.56 & 	$-2.19_{-0.50}^{+0.50}$ \\
N158 & 43.09 & 22.91 & 1.45$^{+ 0.05}_{- 1.17}$ & 2 & 45.05 & 44.25 & 45.37 &   	44.72 & 	$-1.43_{-0.50}^{+0.50}$ \\
N160 & 43.10 & 23.08 & 0.36$^{+ 0.15}_{- 0.12}$ & 4 & 44.83 & 44.25 & 45.25 &   	44.72 & 	$-2.57_{-0.54}^{+0.53}$ \\
N163 & 43.62 & 23.90 & 0.31$^{+ 0.30}_{- 0.30}$ & 3 & 45.02 & 44.78 & 45.00 &   	45.34 & 	$-2.03_{-1.09}^{+0.64}$ \\
N164 & 44.49 & 23.61 & 0.47$^{+ 0.14}_{- 0.11}$ & 4 & 45.11 & 45.65 & 44.99 &   	46.39 & 	$-1.67_{-0.50}^{+0.50}$ \\
N170 & 42.50 & 22.43 & 0.54$^{+ 0.03}_{- 0.03}$ & 3 & 44.44 & 43.66 & 44.76 &   	44.01 & 	$-1.16_{-0.52}^{+0.53}$ \\
N174 & 43.40 & 19.32 & 0.01$^{+ 0.04}_{- 0.00}$ & 2 & 45.09 & 44.56 & 45.43 &   	45.08 & 	$-2.32_{-0.55}^{+0.54}$ \\
N177 & 42.59 & 21.74 & 0.38$^{+ 0.01}_{- 0.37}$ & 3 & 44.86 & 43.75 & 45.31 &   	44.12 & 	$-2.40_{-0.53}^{+0.52}$ \\
N187 & 42.56 & 23.27 & 0.61$^{+ 0.02}_{- 0.60}$ & 4 & 44.74 & 43.72 & 44.99 &   	44.08 & 	$-0.58_{-0.54}^{+0.55}$ \\
N194 & 42.63 & 19.95 & 0.62$^{+ 0.30}_{- 0.08}$ & 4 & 43.89 & 43.79 & 44.37 &   	44.17 & 	$-2.26_{-0.50}^{+0.50}$ \\
N201 & 43.10 & 23.69 & 1.02$^{+ 0.03}_{- 1.01}$ & 2 & 44.71 & 44.26 & 44.79 &   	44.73 & 	$-2.48_{-0.53}^{+0.52}$ \\
N205 & 43.29 & 21.69 & 0.34$^{+ 0.05}_{- 0.06}$ & 4 & 44.92 & 44.45 & 45.04 &   	44.96 & 	$-2.23_{-0.51}^{+0.51}$ \\
N217 & 42.79 & 23.70 & 1.60$^{+ 0.05}_{- 1.59}$ & 4 & 44.57 & 43.94 & 45.00 &   	44.35 & 	$-1.97_{-0.50}^{+0.51}$ \\
N222 & 42.79 & 21.35 & 0.25$^{+ 0.30}_{- 0.24}$ & 3 & 44.95 & 43.95 & 45.36 &   	44.35 & 	$-2.28_{-0.52}^{+0.52}$ \\
N240 & 44.07 & 22.63 & 0.99$^{+ 0.11}_{- 0.22}$ & 2 & 45.22 & 45.23 & 45.41 &   	45.89 & 	$-2.43_{-0.56}^{+0.55}$ \\
N242 & 42.66 & 23.42 & 5.00$^{+ 0.16}_{- 0.16}$ & 4 & 44.29 & 43.82 & 44.50 &   	44.20 & 	$-1.82_{-0.50}^{+0.50}$ \\
N261 & 42.55 & 23.49 & 0.52$^{+ 0.02}_{- 0.51}$ & 4 & 44.58 & 43.71 & 44.80 &   	44.06 & 	$-2.59_{-0.53}^{+0.53}$ \\
N262 & 42.90 & 20.18 & 1.79$^{+ 0.06}_{- 1.77}$ & 1 & 44.21 & 44.06 & 44.74 &   	44.49 & 	$-2.43_{-0.50}^{+0.50}$ \\
N266 & 43.49 & 23.61 & 0.62$^{+ 0.08}_{- 0.06}$ & 4 & 44.95 & 44.65 & 44.90 &   	45.19 & 	$-2.08_{-0.51}^{+0.51}$ \\
N278 & 43.00 & 23.29 & 0.80$^{+ 0.30}_{- 0.14}$ & 1 & 44.72 & 44.16 & 44.72 &   	44.60 & 	$-1.81_{-0.50}^{+0.50}$ \\
N286 & 42.73 & 21.68 & 1.31$^{+ 0.04}_{- 1.30}$ & 2 & 44.34 & 43.89 & 45.10 &   	44.28 & 	$-2.91_{-0.52}^{+0.52}$ \\
N304 & 43.20 & 20.94 & 0.17$^{+ 0.15}_{- 0.07}$ & 3 & 44.62 & 44.35 & 45.07 &   	44.84 & 	$-2.57_{-0.53}^{+0.52}$ \\
N309 & 43.04 & 21.30 & 0.88$^{+ 0.30}_{- 0.54}$ & 1 & 44.59 & 44.20 & 45.04 &   	44.66 & 	$-2.46_{-0.51}^{+0.51}$ \\
N323\tablenotemark{h} & 43.50 & 21.97 & 0.19$^{+ 0.16}_{- 0.09}$ & 4 & 44.90 & 44.66 & 45.18 &   	45.21 & 	$-1.79_{-0.55}^{+0.55}$ \\
N349 & 42.94 & 21.84 & 0.55$^{+ 0.02}_{- 0.01}$ & 4 & 44.44 & 44.09 & 44.61 &   	44.53 & 	$-2.24_{-0.50}^{+0.50}$ \\
N352 & 42.73 & 23.13 & 5.00$^{+ 0.00}_{- 0.16}$ & 3 & 44.70 & 43.89 & 45.29 &   	44.28 & 	$-2.39_{-0.52}^{+0.51}$ \\
N370 & 43.01 & 20.96 & 0.30$^{+ 0.08}_{- 0.07}$ & 4 & 44.65 & 44.17 & 44.95 &   	44.62 & 	$-2.38_{-0.51}^{+0.51}$ \\
N373 & 42.63 & 22.77 & 1.52$^{+ 0.05}_{- 1.51}$ & 4 & 44.11 & 43.79 & 44.36 &   	44.16 & 	$-2.09_{-0.50}^{+0.51}$ \\
N384 & 43.64 & 23.39 & 2.38$^{+ 0.08}_{- 1.63}$ & 2 & 44.93 & 44.80 & 45.15 &   	45.37 & 	$-1.90_{-0.50}^{+0.50}$ \\
N390 & 44.13 & 22.35 & 1.39$^{+ 0.05}_{- 1.37}$ & 1 & 45.66 & 45.29 & 45.59 &   	45.96 & 	$-0.82_{-0.51}^{+0.50}$ \\
N402 & 42.81 & 23.20 & 1.24$^{+ 0.17}_{- 0.18}$ & 4 & 44.47 & 43.97 & 44.64 &   	44.37 & 	$-0.46_{-0.58}^{+0.59}$ \\
N405 & 43.24 & 23.41 & 1.40$^{+ 0.05}_{- 1.34}$ & 3 & 44.83 & 44.40 & 45.25 &   	44.89 & 	$-2.54_{-0.53}^{+0.53}$ \\
N437 & 42.83 & 21.36 & 0.58$^{+ 0.03}_{- 0.11}$ & 4 & 44.62 & 43.99 & 45.31 &   	44.40 & 	$-2.40_{-0.51}^{+0.51}$ \\
N442 & 43.16 & 23.89 & 0.13$^{+ 0.04}_{- 0.02}$ & 4 & 44.97 & 44.32 & 44.85 &   	44.79 & 	$-1.41_{-0.51}^{+0.50}$ \\
N448 & 43.47 & 23.16 & 1.39$^{+ 0.05}_{- 0.97}$ & 1 & 44.89 & 44.62 & 45.09 &   	45.16 & 	$-1.91_{-0.50}^{+0.50}$ \\
N451\tablenotemark{h} & 43.96 & 21.06 & 0.01$^{+ 0.30}_{- 0.00}$ & 3 & 45.53 & 45.11 & 45.53 &   	45.75 & 	$0.29_{-0.60}^{+0.63}$ \\
N471 & 42.84 & 22.90 & 0.96$^{+ 0.19}_{- 0.16}$ & 3 & 44.90 & 44.00 & 45.58 &   	44.41 & 	$-2.73_{-0.55}^{+0.55}$ \\
N473\tablenotemark{h} & 43.28 & 18.23 & 0.13$^{+ 0.01}_{- 0.00}$ & 3 & 44.60 & 44.44 & 45.05 &   	44.94 & 	$-2.33_{-0.71}^{+0.69}$ \\
S 44 & 43.38 & 22.87 & 0.77$^{+ 0.10}_{- 0.11}$ & 3 & 44.69 & 44.54 & 45.00 &	45.06 & 	$-2.11_{-0.50}^{+0.50}$ \\
S 84 & 42.72 & 22.42 & 0.98$^{+ 0.13}_{- 0.21}$ & 2 & 44.72 & 43.87 & 45.06 &	44.26 & 	$-2.61_{-0.53}^{+0.53}$ \\
S 88 & 43.49 & 22.42 & 0.50$^{+ 0.01}_{- 0.00}$ & 4 & 44.94 & 44.65 & 45.21 &	45.19 & 	$-2.70_{-0.56}^{+0.56}$ \\
S 91 & 43.08 & 23.57 & 0.68$^{+ 0.17}_{- 0.22}$ & 3 & 44.72 & 44.24 & 44.96 &	44.70 & 	$-2.02_{-0.50}^{+0.50}$ \\
S103\tablenotemark{h} & 42.90 & 19.98 & 0.95$^{+ 0.03}_{- 0.94}$ & 2 & 44.35 & 44.06 & 44.92 &   	44.48 & 	$-2.40_{-0.50}^{+0.50}$ \\
S117 & 43.12 & 20.46 & 0.01$^{+ 0.00}_{- 0.01}$ & 3 & 44.58 & 44.28 & 44.78 &   	44.75 & 	$-1.60_{-0.50}^{+0.50}$ \\
S118 & 44.23 & 23.30 & 1.31$^{+ 0.30}_{- 1.30}$ & 2 & 45.26 & 45.39 & 45.26 &   	46.08 & 	$>-0.93$ \\
S134 & 43.33 & 23.26 & 0.96$^{+ 0.26}_{- 0.16}$ & 2 & 44.52 & 44.48 & 44.47 &   	44.99 & 	$-1.93_{-0.50}^{+0.50}$ \\
S137 & 42.92 & 23.38 & 0.44$^{+ 0.14}_{- 0.15}$ & 3 & 44.63 & 44.08 & 44.80 &   	44.51 & 	$-2.17_{-0.51}^{+0.51}$ \\
S139 & 42.84 & 23.71 & 0.10$^{+ 0.30}_{- 0.09}$ & 4 & 44.96 & 43.99 & 44.98 &   	44.41 & 	$-1.84_{-0.54}^{+0.54}$ \\
S151 & 42.50 & 21.76 & 1.25$^{+ 0.30}_{- 0.98}$ & 2 & 44.53 & 43.66 & 44.99 &   	44.01 & 	$>-1.56$ \\
S155 & 42.73 & 21.22 & 0.41$^{+ 0.09}_{- 0.11}$ & 3 & 44.16 & 43.89 & 44.76 &   	44.28 & 	$-2.46_{-0.50}^{+0.50}$ \\
S156 & 43.28 & 22.65 & 0.43$^{+ 0.30}_{- 0.29}$ & 4 & 44.72 & 44.43 & 44.91 &   	44.94 & 	$-2.25_{-0.52}^{+0.51}$ \\
S159 & 43.57 & 21.76 & 1.22$^{+ 0.04}_{- 1.10}$ & 1 & 44.63 & 44.73 & 44.75 &   	45.29 & 	$-2.36_{-0.51}^{+0.51}$ \\
S160 & 43.02 & 23.19 & 0.28$^{+ 0.30}_{- 0.20}$ & 4 & 44.58 & 44.18 & 44.16 &   	44.63 & 	$-1.90_{-0.97}^{+0.64}$ \\
S161 & 43.19 & 22.86 & 0.47$^{+ 0.05}_{- 0.07}$ & 3 & 44.69 & 44.35 & 44.84 &   	44.83 & 	$-2.40_{-0.52}^{+0.51}$ \\
S162 & 42.69 & 22.69 & 0.22$^{+ 0.01}_{- 0.21}$ & 4 & 44.55 & 43.85 & 44.49 &   	44.24 & 	$-2.47_{-0.53}^{+0.53}$ \\
S164 & 43.53 & 23.56 & 1.03$^{+ 0.03}_{- 1.02}$ & 2 & 45.04 & 44.69 & 45.29 &   	45.24 & 	$-2.42_{-0.54}^{+0.54}$ \\
S171 & 42.55 & 21.32 & 0.53$^{+ 0.02}_{- 0.02}$ & 2 & 44.44 & 43.71 & 44.78 &   	44.06 & 	$>-1.41$ \\
S176 & 43.01 & 22.35 & 0.01$^{+ 0.30}_{- 0.00}$ & 3 & 45.01 & 44.16 & 44.87 &   	44.61 & 	$-1.58_{-0.54}^{+0.52}$ \\
S179\tablenotemark{h} & 43.38 & 22.85 & 0.90$^{+ 0.03}_{- 0.89}$ & 2 & 44.87 & 44.54 & 44.82 &   	45.06 & 	$-1.93_{-0.50}^{+0.50}$ \\
S184 & 42.86 & 21.16 & 0.01$^{+ 0.00}_{- 0.01}$ & 3 & 44.74 & 44.02 & 45.16 &   	44.44 & 	$-2.26_{-0.55}^{+0.54}$ \\
S193\tablenotemark{h} & 43.29 & 21.05 & 0.53$^{+ 0.30}_{- 0.18}$ & 2 & 44.90 & 44.45 & 45.18 &   	44.95 & 	$>-1.00$ \\
S200 & 42.92 & 21.91 & 0.35$^{+ 0.01}_{- 0.02}$ & 4 & 44.77 & 44.08 & 44.83 &   	44.51 & 	$-1.32_{-0.51}^{+0.51}$ \\
S203 & 42.75 & 22.81 & 0.44$^{+ 0.06}_{- 0.10}$ & 4 & 44.83 & 43.90 & 45.35 &   	44.30 & 	$-3.01_{-0.57}^{+0.57}$ \\
S214 & 43.16 & 21.70 & 0.42$^{+ 0.03}_{- 0.07}$ & 3 & 44.68 & 44.32 & 44.81 &   	44.80 & 	$-2.14_{-0.51}^{+0.50}$ \\
S220 & 43.21 & 23.23 & 0.64$^{+ 0.06}_{- 0.06}$ & 3 & 44.56 & 44.37 & 44.71 &   	44.86 & 	$-2.22_{-0.50}^{+0.50}$ \\
S227 & 43.32 & 21.46 & 0.49$^{+ 0.01}_{- 0.01}$ & 2 & 44.38 & 44.47 & 44.52 &   	44.98 & 	$-2.33_{-0.51}^{+0.50}$ \\
S229 & 42.98 & 20.07 & 0.06$^{+ 0.03}_{- 0.01}$ & 2 & 44.36 & 44.14 & 44.53 &   	44.58 & 	$-1.26_{-0.52}^{+0.53}$ \\
S249 & 43.36 & 23.74 & 0.47$^{+ 0.12}_{- 0.15}$ & 4 & 44.89 & 44.52 & 44.88 &   	45.04 & 	$-1.80_{-0.50}^{+0.50}$ \\
S263\tablenotemark{h} & 43.47 & 22.77 & 5.00$^{+ 0.16}_{- 0.16}$ & 1 & 44.31 & 44.63 & 44.70 &   	45.17 & 	$-2.67_{-0.51}^{+0.51}$ \\
S271 & 43.29 & 23.11 & 0.54$^{+ 0.30}_{- 0.53}$ & 2 & 44.97 & 44.45 & 45.04 &   	44.96 & 	$-2.19_{-0.52}^{+0.52}$ \\
S273 & 42.41 & 21.37 & 0.20$^{+ 0.30}_{- 0.16}$ & 2 & 44.33 & 43.57 & 44.79 &   	43.90 & 	$-2.39_{-0.50}^{+0.50}$ \\
S276 & 42.65 & 22.65 & 0.01$^{+ 0.10}_{- 0.00}$ & 3 & 44.97 & 43.81 & 45.13 &   	44.19 & 	$-2.40_{-0.62}^{+0.57}$ \\
S286\tablenotemark{h} & 43.74 & 20.85 & 0.30$^{+ 0.30}_{- 0.20}$ & 2 & 45.27 & 44.89 & 45.29 &   	45.48 & 	$-0.68_{-0.51}^{+0.51}$ \\
S293 & 42.56 & 19.49 & 5.00$^{+ 0.00}_{- 0.16}$ & 2 & 44.28 & 43.71 & 45.06 &   	44.07 & 	$-3.18_{-0.55}^{+0.54}$ \\
S300 & 42.74 & 21.50 & 0.66$^{+ 0.30}_{- 0.65}$ & 4 & 44.56 & 43.90 & 44.68 &   	44.30 & 	$-1.38_{-0.51}^{+0.51}$ \\
\enddata
\tablenotetext{a}{Method 1 in Section \ref{lbol_compare}; described in detail in Section \ref{dust_lum}.}
\tablenotetext{b}{Method 2 in Section \ref{lbol_compare}; described in detail in Section \ref{KX}.}
\tablenotetext{c}{Method 3 in Section \ref{lbol_compare}.}
\tablenotetext{d}{Method 4 in Section \ref{lbol_compare}; from \citet{treister09}.}
\tablenotetext{e}{All luminosities are in \emph{cgs} (erg s$^{-1}$).}
\tablenotetext{f}{Gas column density in cm$^{-2}$.}
\tablenotetext{g}{Calculated using \lbol\ from Method 1.}
\tablenotetext{h}{Broad-Line AGN \citep{cowie03,wirth04,szokoly04}.}
\end{deluxetable}

%\renewcommand{\thefootnote}{\arabic{footnote}}

%%%%%%%%%%%%%%%%%%%%%%%%%% End Table 2 %%%%%%%%%%%%%%%%%%%%%%%%%%%%

\end{document}